%% file: KepImn.tex
\documentclass[useAMS,usenatbib,usegraphicx]{mn2e}

\input{defs}

\newcommand{\figref}[1]{Fig.~\ref{#1}}

\title[First {\em Kepler} results on compact pulsators]
{First {\em Kepler} results on compact pulsators\\
I. Survey target selection and the first pulsators }

\author[R.~H.~{\O}stensen et al.]
       {R.~H.~{\O}stensen,$^1$\thanks{E-mail: roy@ster.kuleuven.be}
        R.~Silvotti,$^2$
        S.~Charpinet,$^3$
        R.~Oreiro,$^{1,4}$
        G.~Handler,$^5$\newauthor
        E.~M.~Green,$^6$
        S.~Bloemen,$^1$
        U.~Heber,$^{7}$
        B.~T.~G\"ansicke,$^8$
        T.~R.~Marsh,$^8$\newauthor
        D.~W.~Kurtz,$^{9}$
        J.~H.~Telting,$^{10}$
        M.~D.~Reed,$^{11}$
        S.~D.~Kawaler,$^{12}$
        C.~Aerts,$^{1,13}$\newauthor
        C.~Rodr{\'{\i}}guez-L{\'o}pez,$^{3,14}$
        M.~Vu\v{c}kovi\'{c},$^{1,15}$
        T.~A.~Ottosen,$^{10,16}$
        T.~Liimets,$^{10,17}$\newauthor
        A.~C.~Quint,$^{11}$
        V.~Van Grootel,$^{3}$
        S.~K.~Randall,$^{18}$
        R.~L.~Gilliland,$^{19}$
        H.~Kjeldsen,$^{15}$ \newauthor
        J.~Christensen-Dalsgaard,$^{15}$
        W.~J.~Borucki,$^{20}$
        D.~Koch$^{20}$ and
        E.~V.~Quintana$^{21}$\\
        $^1$ Instituut voor Sterrenkunde, K.U.~Leuven, Celestijnenlaan 200D, 3001 Leuven, Belgium \\
        $^2$ INAF-Osservatorio Astronomico di Torino, Strada dell'Osservatorio 20, 10025 Pino Torinese, Italy \\
        $^3$ Laboratoire d'Astrophysique de Toulouse-Tarbes, Univ.~de Toulouse,
 14 Av.~Edouard Belin, Toulouse 31400, France \\
        $^4$ Instituto de Astrof\'isica de Andaluc\'ia, Glorieta de la Astronom\'\i
a s/n, 18008 Granada, Spain \\
        $^5$ Institut f\"ur Astronomie, Universit\"at Wien, T\"urkenschanzstrasse 17, 1180 Wien, Austria \\
        $^6$ Steward Observatory, University of Arizona, 933 N.~Cherry Ave., Tucson, AZ 85721, USA \\
        $^{7}$ Dr. Karl Remeis-Observatory \& ECAP, Astronomical Inst.,
FAU Erlangen-Nuremberg, Sternwartstr.~7, 96049 Bamberg, Germany \\
        $^8$ Department of Physics, University of Warwick, Coventry CV4 7AL, UK \\
        $^{9}$ Jeremiah Horrocks Institute of Astrophysics, University of Central Lancashire, Preston PR1 2HE, UK \\
        $^{10}$ Nordic Optical Telescope, 38700 Santa Cruz de La Palma, Spain \\
        $^{11}$ Department of Physics, Astronomy, and Materials Science, Missouri State University, Springfield, MO 65804, USA \\
        $^{12}$ Department of Physics and Astronomy, Iowa State University, Ames, IA 50011, USA \\
        $^{13}$ Department of Astrophysics, IMAPP, Radboud University
   Nijmegen, 6500 GL Nijmegen, The Netherlands \\
        $^{14}$ Departemento de F\'\i sica Aplicada, Univ. de Vigo, Campus Lagoas-Marcosende s/n, 36310 Vigo, Spain \\
        $^{15}$ European Southern Observatory, Alonso de C\'ordova 3107,
               Vitacura, Casilla 19001, Santiago, Chile\\
        $^{16}$ Department of Physics and Astronomy, Aarhus University, 8000 Aarhus C, Denmark \\
        $^{17}$ Tartu Observatoorium, T\~oravere, 61602, Estonia \\
        $^{18}$ ESO, Karl-Schwarzschild-Str.~2, 85748 Garching bei M\"unchen, Germany \\
        $^{19}$ Space Telescope Science Institute, 3700 San Martin Drive, Baltimore, MD 21218, USA \\
        $^{20}$ NASA Ames Research Center, MS 244-30, Moffett Field, CA 94035, USA \\
        $^{21}$ SETI Institute, NASA Ames Research Center, MS 244-30, Moffett Field, CA 94035, USA
}

\begin{document}

\date{Released 2010 Xxxxx XX}
\pagerange{\pageref{firstpage}--\pageref{lastpage}} \pubyear{2010}
\maketitle
\label{firstpage}

\begin{abstract}
We present results from the first two quarters of a survey to search for
pulsations in compact stellar objects with the \kep\ spacecraft.
The survey sample and the various methods applied in its compilation
are described, and spectroscopic observations are presented to
separate the objects into accurate classes.
From the \kep\ photometry we clearly identify nine compact pulsators, and
a number of interesting binary stars.
Of the pulsators, one shows the strong, rapid pulsations typical for
a V361\,Hya type sdB variable (sdBV), seven show long-period
pulsations characteristic of V1093\,Her type sdBVs, and one shows
low-amplitude pulsations with both short and long periods.
We derive effective temperatures and surface gravities for all the
subdwarf\,B stars in the sample and demonstrate that below the boundary
region where hybrid sdB pulsators are found,
all our targets are pulsating. For the stars hotter than this boundary
temperature a low fraction of strong pulsators ($<$10 per cent) is confirmed.
Interestingly, the short-period pulsator also shows a low-amplitude mode in
the long-period region, and several of the V1093\,Her pulsators show low
amplitude modes in the short-period region, indicating that hybrid
behaviour may be common in these stars, also outside the boundary
temperature region where hybrid pulsators have hitherto been found.
\end{abstract}

\begin{keywords}
   subdwarfs -- white dwarfs -- surveys --
   stars: oscillations --
   binaries: close --
   Kepler
\end{keywords}

\section{Introduction}

Among the different classes of pulsating stars, the various subclasses
of hot subdwarfs and white dwarfs show comparable observational
characteristics, with pulsation periods that can be as short as
$\sim$1\,min, due to the compact nature of these objects
\citep{asteroseismology}.  The term {\em compact pulsators},
used in this series of articles, refers to all
these compact oscillating stars as a group.
However, in the first half of the survey phase of the \kepmi,
no strongly pulsating white dwarfs (WDs) were found, and for this
reason we will concentrate primarily on the pulsating subdwarf\,B (sdB)
stars, of which we have identified nine clear cases.

Most sdBs are extreme horizontal branch (EHB) stars, which identifies them
as core helium-burning post-RGB stars with a mass close to the helium flash
mass of 0.5\,\msol, and hydrogen envelopes too thin to sustain
shell burning.
In order for them to reach this configuration, an extraordinary mechanism
is required to remove almost the entire envelope when they are
close to the tip of the RGB.
Three distinct scenarios involving binary interactions have been identified
and found to be sufficient to establish the EHB stars as the cause of the
UV-upturn phenomenon seen in elliptical galaxies
\citep[see][for a review]{podsi08}.
But many questions remain, both with respect to the contributions from
the different binary channels and especially regarding the creation of single
sdB stars.

Pulsations in sdB stars were discovered by \citet{kilkenny97} and they are now
known as sdBV stars or V361\,Hya stars, after the prototype
(EC\,14026--2647). These stars are multiperiodic pulsators
with very short periods, typically 2\,--\,3\,min, and have effective
temperatures between 28\,000 and 37\,000\,K and surface gravities
typically between \logg\,[cgs]\,=\,5.5 and 6.1 \citep{sdbnot}.
Some years after the discovery of short-period pulsations,
\citet{green03} reported pulsations with much longer
periods ($\sim$1\,h) in sdB stars with temperatures below
$\sim$30\,000\,K\footnote{Note that this limit is based on temperatures that
were derived by fits to metal-free NLTE models.
Metal blanketed LTE models such as those used here
typically produce effective temperatures that are cooler by 5\,--\,10 per cent.}. 
These stars are often referred to as long-period sdBVs or V1093\,Her stars
after the prototype.
More recently, \citet{schuh06} found a star on the boundary between
the V361\,Hya and V1093\,Her stars that exhibits simultaneous short- and
long-period pulsations. Such hybrid pulsators are often referred to as
DW\,Lyn stars after the prototype.
Now, four stars in this temperature boundary region
have been found with both types of pulsations. In addition to DW\,Lyn they
are Balloon\,090100001 \citep{oreiro04,oreiro05},
V391\,Peg \citep{ostensen01a,lutz09}, and
RAT\,0455+1305 \citep{ramsay06,baran10}.
With the exception of V338\,Ser, all pulsating sdB stars reside on or very
close to the canonical Extreme Horizontal Branch
(see \citealt{ostensen09}, \citealt{charpinet09} and \citealt{heber09}
for recent detailed reviews of
asteroseismology of EHB stars and hot subdwarfs in general).

Due to the recent discoveries of various subclasses of pulsators in
the sdB family, a common nomenclature has yet to emerge. 
\citet{kilkenny10b} have recently proposed to label the different
type of sdBV stars with subscripts, so that short-period pulsators
are labeled sdBV$_{\rm r}$ and long-period pulsators sdBV$_{\rm s}$
(for {\em rapid} and {\em slow}) and pulsators with hybrid behaviour as
sdBV$_{\rm rs}$.
However, as we will learn from the data presented here, it may be too
early yet to settle this issue, since many more stars than originally
anticipated appear to show hybrid behaviour at low amplitudes.
This means that we have to face the possibility
that all sdBVs are hybrids at some level.
For this reason we will stick to using the variable star classes as
designators for the dominant pulsational behaviour, and reserve the
DW\,Lyn type designation for stars that show clear and unambiguous
pulsations at both short and long periods, while we refer to stars
that are predominantly V361\,Hya (or V1093\,Her) stars but with low
level pulsations of the other type as being of the predominant type
with hybrid behaviour.  

Non-adiabatic pulsation calculations
show that the V361\,Hya stars exhibit low $n$, low $\ell$ pressure
modes excited by the $\kappa$ mechanism due to a $Z$-bump in the envelope
\citep{charpinet97}, and the V1093\,Her stars can be understood in
terms of gravity modes of high $n$ excited by the same $\kappa$ mechanism
\citep{fontaine03}. The pulsations are facilitated by radiative levitation,
which ensures that sufficient iron-group elements are present to cause
a Z-bump in the driving region.

Our various surveys have found that only one in ten sdBs in the hot end of
the instability region pulsates with short-period oscillations \citep{sdbnot},
while long-period pulsations appear in at least 75 per cent of all stars in the
cool end of the instability region \citep{green03},
as determined from ground-based observations.
The amplitudes for the V1093\,Her stars are so low that it has been speculated
that most sdB stars cooler than the boundary between the V361\,Hya and the
V1093\,Her stars are pulsators, but with such low amplitudes that they
are undetectable without extraordinary efforts from ground-based observatories.

It is an unsolved mystery why so few of the hot stars oscillate. Diffusion
calculations have shown that sufficient iron to excite pulsations builds up in
the driving region of sdB stars after only 10$^6$ years \citep{fontaine06},
which is just 1 per cent of the EHB lifetime, implying that $\sim$99 per cent should pulsate.
Most short-period sdBVs that have been monitored regularly over the past
decade have shown pulsations with strongly variable amplitudes
\citep{kilkenny10a}. Some pulsation frequencies have even been reported to
disappear completely. It is therefore possible that many of the stars that
have been surveyed and found not to show variability, may in fact
just be passing through an intermittent, quiet phase.
With the continuous monitoring provided by \kep, we will
be able to quantify both fast and slow pulsations with much lower amplitudes than can be
reached from ground-based observations, and determine if pulsation
episodes occur.

The \kep\ spacecraft was launched in March of 2009, with the primary aim
to find Earth sized planets using the transit method \citep{borucki10}.
In order to have a high probability
of finding such planets, the spacecraft is designed to continuously monitor the
brightness of a massive number of stars with micromagnitude precision.
The \kep\ Field of View (FoV) covers 105 square degrees in
Cygnus and Lyra (about 10\,--\,20 degrees from the galactic plane) using 42 CCDs.
As a byproduct, high quality photometric data of pulsating stars in the same field
are obtained, an incredibly valuable input for asteroseismology
studies \citep{gilliland10a}, and a unique opportunity for asteroseismology studies of V1093\,Her
stars in particular, since adequate observations are nearly impossible to obtain
from the ground \citep{randall06}.

The first four quarters of the \kepmi\ were dedicated to a survey phase,
for the asteroseismology subset of targets.
A substantial number of target buffers for short-cadence observations have been made
available to the \kep\ {\em Asteroseismic Science Consortium} ({\sc kasc}) during this
survey phase. After the survey phase, {\sc kasc} observations will only be possible on
specific targets selected from the survey sample.
The primary goal of the survey phase is therefore to identify the most interesting
pulsators in the sample. These objects can then be followed throughout the
remaining years of the mission, with goals including:
i)~Detecting low amplitude ($<$100\,ppm) and high degree ($\ell >2$) modes,
normally not visible from the ground;
ii)~Measuring stellar global parameters with unprecedented accuracy
(mass, rotation, envelope thickness, and surface gravity);
iii)~Improving the physical understanding of these stars (differential rotation,
core C/O ratio, neutrino cooling and crystallisation in the WDs);
iv)~Studying amplitude variability and non-linear effects;
v)~Via the O--C diagram, measure the rate of period change, 
determine the evolutionary status of the star, and search for low-mass
companions \citep[brown dwarfs or planets,][]{silvotti07}
with masses down to $\sim$0.1\,M$_{\rm{J}}$.

This paper is the first in a series on compact pulsators in the
{\em Kepler} field, and describes the methods with which the candidates
were selected, and provides classifications and noise limits on the targets
in the first half of the survey sample. 
The first analysis of a V361\,Hya star in the \kep\ field are presented
by \citet[][Paper {\sc ii}]{kawaler10a},
the first results on V1093\,Her and DW\,Lyn pulsators
are presented by \citet[][Paper {\sc iii}]{reed10a},
and a detailed asteroseismic analysis of \kep\ data on one of these DW\,Lyn
pulsators are given by \citet[][Paper {\sc iv}]{VanGrootel10}.
Results on two V1093\,Her pulsators in sdB+dM reflection binaries are
discussed in \citet[][Paper {\sc v}]{kawaler10b}.

\begin{table*}
\caption[]{Compact pulsator candidates observed with \kep.}
\label{tbl:targets}
\centering
\begin{tabular}{rllllllll} \hline
\tabkic  & Name          & Run  & RA(J2000) & Dec(J2000) & \mkep &\corr& Sample & Class \\ \hline
 1868650 & KBS\,13       & Q1   & 19:26:09.4 & +37:20:09  & 13.45 &0.158& a & sdB+dM \\
 2297488 & J19208+3741   & Q1   & 19:20:49.9 & +37:41:39  & 17.18 &0.833& c & sdO+F/G \\
 2692915 & J19033+3755   & Q2.3 & 19:03:22.7 & +37:55:30  & 17.54 &0.758& d & DO \\
 2697388 & J19091+3756   & Q2.3 & 19:09:07.1 & +37:56:14  & 15.39 &0.149& cf & sdBV \\
 2991276 & J19271+3810   & Q2.1 & 19:27:09.1 & +38:10:26  & 17.42 &0.971& c & sdB \\
 2991403 & J19272+3808   & Q1   & 19:27:15.9 & +38:08:08  & 17.14 &0.601& c & sdBV+dM \\
 3427482 & J19053+3831   & Q1   & 19:05:22.5 & +38:31:33  & 17.31 &0.891& c & DA \\
 3527617 & J19034+3841   & Q2.2 & 19:03:24.9 & +38:41:24  & 17.54 &0.554& c & He-sdOB \\
 3527751 & J19036+3836   & Q2.3 & 19:03:37.0 & +38:36:13  & 14.86 &0.081& cf & sdBV \\
 3729024 & J19014+3852   & Q2.2 & 19:01:25.8 & +38:52:33  & 17.63 &0.611& c & sdB \\
 4244427 & J19032+3923   & Q2.1 & 19:03:17.6 & +39:23:14  & 17.35 &0.781& c & sdB \\
 4829241 & J19194+3958   & Q1   & 19:19:27.7 & +39:58:39  & 15.83 &0.435& e$^\dag$ & DA \\
 5342213 & J18568+4032   & Q2.2 & 18:56:49.5 & +40:32:51  & 17.68 &0.362& c & sdOB \\
 5775128 & 2M1905+4101   & Q2.3 & 19:05:01.0 & +41:01:53  & 13.54 &0.023& bf & B \\
 5807616 & KPD\,1943+4058& Q2.3 & 19:45:25.5 & +41:05:34  & 15.02 &0.332& a & sdBV \\
 5942605 & FBS\,1858+411 & Q2.1 & 19:00:27.8 & +41:15:04  & 14.08 &0.088& a & B \\
 6188286 & J19028+4134   & Q2.3 & 19:02:49.9 & +41:34:58  & 14.21 &0.028& f & sdOB \\
 6669882 & J18557+4207   & Q2.3 & 18:55:46.0 & +42:07:04  & 17.94 &0.468& d & DA \\
 6848529 & \bd           & Q0   & 19:07:40.5 & +42:18:22  & 10.73 &0.003& a$^\ddag$ & sdB \\
 6862653 & J19267+4219   & Q2.3 & 19:26:46.0 & +42:19:34  & 18.19 & ... & d & DB \\
 7353409 & 2M1915+4256   & Q2.2 & 19:15:18.9 & +42:56:13  & 14.68 &0.091& b & sdO \\
 7434250 & J19135+4302   & Q2.3 & 19:13:35.6 & +43:02:56  & 15.47 &0.287& d & sdB \\
 7664467 & J18561+4319   & Q2.3 & 18:56:07.1 & +43:19:19  & 16.45 &0.879& f & sdBV \\
 7755741 & 2M1931+4324   & Q1   & 19:31:08.9 & +43:24:58  & 13.75 &0.112& b & sdO \\
 7975824 & KPD\,1946+4340& Q1   & 19:47:42.9 & +43:47:31  & 14.65 &0.106& a$^\ddag$ & sdOB+WD \\
 8022110 & J19179+4350   & Q2.3 & 19:17:58.7 & +43:50:21  & 16.55 &0.316& f & sdB \\
 8077281 & J18502+4358   & Q2.3 & 18:50:16.9 & +43:58:28  & 16.57 &0.296& f$^\dag$ & B \\
 8142623 & J18427+4404   & Q1   & 18:42:42.4 & +44:04:05  & 17.30 &0.388& de$^\dag$ & sdB \\
 8496196 & J19288+4430   & Q2.3 & 19:28:51.1 & +44:30:13  & 16.45 &0.827& f & sdOB \\
 8619526 & J19195+4445   & Q1   & 19:19:30.5 & +44:45:43  & 15.84 &0.256& f$^\dag$ & PNN \\
 8682822 & J19173+4452   & Q1   & 19:17:20.6 & +44:52:40  & 15.81 &0.473& e$^\dag$ & DA \\
 8751494 & J19241+4459   & Q2.2 & 19:24:10.8 & +44:59:35  & 16.27 &0.566& e$^\dag$ & CV \\
 8889318 & J19328+4510   & Q2.3 & 19:32:50.6 & +45:10:23  & 17.17 &0.645& d & sdB \\
 9139775 & J18577+4532   & Q2.3 & 18:57:43.6 & +45:32:19  & 17.92 &0.272& d & DA \\
 9202990 & J18561+4537   & Q2.3 & 18:56:08.0 & +45:37:40  & 15.04 &0.119& d$\ddag$ & CV \\
 9408967 & J19352+4555   & Q2.3 & 19:35:12.2 & +45:55:42  & 17.26 &0.807& d & He-sdOB \\
 9472174 & 2M1938+4603   & Q0   & 19:38:32.6 & +46:03:59  & 12.26 &0.022& b & sdBV+dM \\
 9543660 & 2M1951+4607   & Q1   & 19:51:56.2 & +46:07:51  & 13.77 &0.459& b & sdOB \\
 9569458 & J18477+4616   & Q1   & 18:47:43.6 & +46:16:26  & 17.18 &0.031& d & sdB \\
 9583158 & NSV\,11917    & Q2.1 & 19:19:10.2 & +46:14:51  & 17.32 &0.465& ad & PNN \\
 9822180 & J19100+4640   & Q2.1 & 19:10:00.3 & +46:40:25  & 14.58 &0.215& d & sdO+F/G \\
 9957741 & J19380+4649   & Q2.1 & 19:38:01.6 & +46:49:45  & 16.10 &0.099& f & He-sdOB \\
10130954 & 2M1910+4709   & Q0   & 19:10:23.7 & +47:09:44  & 11.13 &0.001& bf & B \\
10139564 & J19249+4707   & Q2.1 & 19:24:58.2 & +47:07:54  & 16.13 &0.608& f & sdBV \\
10220209 & 2M1945+4714   & Q2.3 & 19:45:33.2 & +47:14:17  & 14.12 &0.090& bf & B \\
10285114 & 2M1944+4721   & Q0   & 19:44:02.7 & +47:21:17  & 11.23 &0.028& bf & B \\
10420021 & J19492+4734   & Q2.2 & 19:49:14.6 & +47:34:46  & 16.20 &0.909& e & DA \\
10482860 & J19458+4739   & Q2.3 & 19:41:34.2 & +49:07:14  & 17.80 &0.640& d & DO \\
10593239 & J19162+4749   & Q2.3 & 19:16:12.2 & +47:49:16  & 15.28 &0.120& d & sdB+F/G \\
10658302 & 2M1915+4754   & Q2.1 & 19:15:07.9 & +47:54:20  & 13.12 &0.148& bf & B \\
10661778 & J19211+4759   & Q2.3 & 19:21:11.1 & +47:59:24  & 17.66 &0.483& d & sdB \\
10670103 & J19346+4758   & Q2.3 & 19:34:39.9 & +47:58:12  & 16.53 &0.450& f & sdBV \\
10982905 & J19405+4827   & Q2.1 & 19:40:32.2 & +48:27:24  & 14.15 &0.154& d & sdB+F/G \\
11032470 & J19327+4834   & Q2.2 & 19:32:45.3 & +48:34:47  & 16.13 &0.680& f & F \\
11179657 & J19023+4850   & Q2.3 & 19:02:21.9 & +48:50:52  & 17.06 &0.129& d & sdBV+dM \\
11357853 & J19415+4907   & Q2.1 & 19:41:34.3 & +49:07:14  & 17.37 &0.632& d & sdOB \\
11454304 & 2M1925+4918   & Q0   & 19:25:24.4 & +49:18:56  & 12.95 &0.007& bf & B \\
11514682 & J19412+4925   & Q2.3 & 19:41:12.4 & +49:25:06  & 15.69 &0.436& d & DAB \\
11717464 & J19369+4953   & Q1   & 19:36:55.8 & +49:53:14  & 15.73 &0.525& f & A \\
11817929 & 2M1935+5002   & Q0   & 19:35:38.4 & +50:02:35  & 10.38 &0.002& bf& B \\
11822535 & WD\,1942+499  & Q2.2 & 19:43:41.8 & +50:05:01  & 14.82 &0.155& a & DA \\
11953267 & 2M1901+5023   & Q2.2 & 19:01:35.2 & +50:23:06  & 13.50 &0.071& bf& B \\
12156549 & J19193+5044   & Q2.3 & 19:19:18.8 & +50:44:04  & 15.89 &0.107& d & DA+dMe \\ \hline
\end{tabular}\\
{\sc Notes.}---\mkep\ is the magnitude in the \kep\ bandpass, \corr\ is the
contamination factor from the \kic. The samples are:
a: Literature, b: \twom, c: SDSS, d: {\em Galex}, e: RPM, f: \kic\ color.
$^\dag$ and $^\ddag$ mark targets with TNG and {\sc not} photometry respectively.
\end{table*}

\section{Survey sample selection}\label{sect:select}

For the survey phase of the \kepmi, three groups submitted proposals
containing candidate hot subdwarf and white dwarf stars. Of the stars included
in these proposals, 142 were accepted into the list of {\sc kasc} survey stars.
Of these, six were observed during the 9.7-day commissioning run, and 57 were observed
during the first four (out of ten) survey months. All 63 stars are listed in
Table~\ref{tbl:targets}, where we also give a name for each target from the literature,
or an abbreviated coordinate designation. The survey month, target coordinates,
\kep\ magnitude (\mkep)\footnote{The \mkep\ magnitudes approximates the passband of
the \kep\ photometer and is computed from $B$ and $V$ magnitudes from the Tycho-2
catalogue. For blue stars the formula is \mkep\,=\,0.344$B$ + 0.656$V$ $-$ 0.032.
}, estimated contamination factor (\corr),
original survey sample, and our final classification
(see Section~\ref{sect:spectro}) are also listed in the table.
Note that the large FoV of \kep\ implies that the resolution is only around 4 arcsec
per pixel \citep{bryson10}. Typically, for our faint
targets, four to ten pixels are summed to capture all the flux, with the number increasing
for targets located towards the edges of the field.
Contamination from nearby stars can be severe, as indicated by the sometimes large
\corr\ values in Table~\ref{tbl:targets}.
Survey months for the \kepmi\ are referred to in the following way; Q0 refers
to the commissioning period, Q1 refers to the first 33.5-d survey cycle
(12 May -– 14 June 2009), Q2 refers to the second quarter of the survey phase (90 days),
subdivided into three monthly cycles.
All objects are referred to by their identifier from the {\em Kepler Input Catalog} (\kic),
but we will refer to many of the stars listed here by their name from the survey
that first identified them as compact stars.
For convenience, both names are listed in the tables in this paper.
For the stars from the SDSS and {\em Galex} surveys, we have
used abbreviated coordinate names throughout our spectroscopic survey,
and these are retained in the tables and discussions here.
When we start discussing the photometric properties of these targets,
we will prefer the \kic\ identifiers, in order to
conform with other works on \kep\ photometry.  Note that even if the table
is sorted by \kic\ identifiers, it is easy to locate
an object by coordinate name, since \kic\ numbers are allocated in order of
increasing declination.

The different methods used to construct this sample are summarised in the
following sections.

\subsection{Known compact stars from the literature}
A few subdwarfs from earlier blue-star surveys were already known to be located
within the \kep\ FoV.
The bright sdB star \bd\ ($B$\,=\,10.4) was known
not to pulsate from the study of \citet{sdbnot}, but included in the survey anyway
since its brightness permits by far the highest S/N ratio of the stars in the sample,
potentially revealing the presence of extremely low level pulsations.
KPD\,1946+4340 and KPD\,1943+4058 were
reported to be hot subdwarf stars in the {\em Kitt Peak Downes} survey \citep{kpd}.
KPD\,1946+4340 was also known to be a binary system with a period of 0.4\,d from the
radial velocity survey of \citet{morales-rueda03}. It was also included in the
variability survey of \citet{sdbnot}, but was not found to vary on short time-scales
above the millimagnitude level.

Also the {\em First Byurakan Survey} for blue stellar objects \citep[FBS,][]{abrahamian90}
contains a few fields
that overlap with the \kep\ FoV. Three stars from the FBS were identified
as likely sdB stars, but none of them has been studied in detail up to now.
The star KBS\,13 was identified in the first {\em Kepler Blue Star} (KBS) survey, and
identified as a reflection effect sdB+dM binary by \citet{for08}.

Two white dwarf stars in the \kep\ field were also known from the literature.
WD\,1942+499 is a DA white dwarf, but with a temperature of 33\,500\,K
\citep{marsh97} it is far too hot to reside in the ZZ\,Ceti instability region.
NSV\,11917 is the planetary nebula nucleus (PNN) of
Abell\,61 \citep{abell66},
and its temperature was estimated to be $\sim$88\,000\,K by \citet{napiwotzki99}.

\subsection{The 2{\sevensize\bf MASS} sample}
A list of 171 candidates was obtained from the \twom\ catalogue \citep{twomass}
with $J < 15.5$, and colors
covering the range of hot subdwarfs, $J-H < 0.01$, $J-K < 0.01$, that were within the FoV
of the \kep\ CCDs and more than five pixels from the edges. They were then observed
spectroscopically with the {\em Bok telescope} on Kitt Peak.
Unfortunately, of all the targets only two new sdB stars and two new sdO stars
were found. Detecting compact stars from \twom\ colours has produced much
higher success rates in high galactic latitude surveys, but since the \kep\
field is located relatively close to the galactic plane the contamination of
main-sequence B
stars is very high down to the limiting magnitude of the \twom\ survey.

Also, one of the stars classified as sdB in the FBS survey appeared in
this sample, but the spectroscopy revealed that it was actually another main
sequence B star.  Pre-launch ground-based photometry of the four hot
subdwarfs revealed that one of them (2M1938+4603)
has a strong reflection effect with grazing eclipses indicating
the presence of a close M-dwarf companion \citep{2m1938}.

\subsection{The \textbfit{Galex} sample}
In order to avoid the contamination of hot main-sequence stars, we searched the available
archives for ultraviolet (UV) photometry.
A reliable photometric characteristic of both hot subdwarfs and white dwarfs is the
strong UV-excess relative to normal stars.
The UV photometry from the {\em Galex} satellite \citep{GALEX} covers large patches of the northern
half of the \kep\ field, and Guest Observer (GO) fields cover further patches
across the field.
In total we identified 27 UV-excess targets from the {\em Galex} GR4 catalogue,
and a further 24 from guest observer fields kindly made available to us.
As expected, selecting targets based on UV photometry turned out to be extremely successful.
Of the twenty stars in the current sample
that were selected from {\em Galex} photometry, all are hot subdwarfs or white dwarfs.

\begin{figure}
\centering
\includegraphics[width=\hsize]{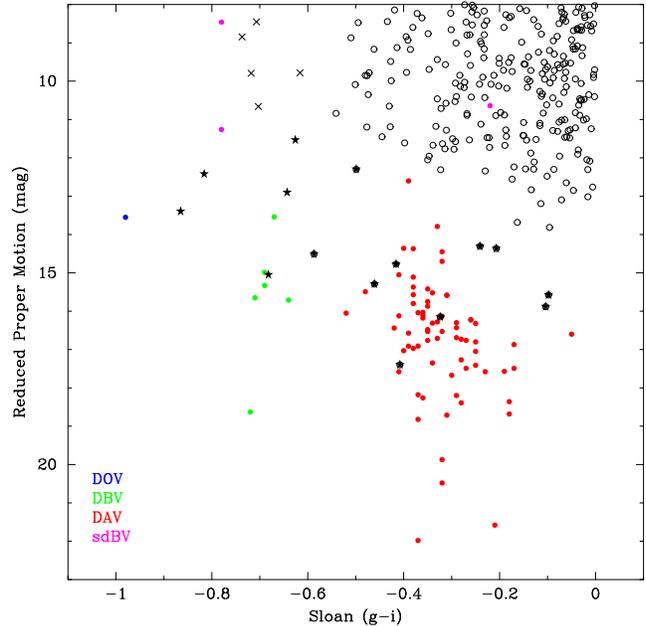}
\caption{The reduced proper motion diagram used to select targets from the \kic.
Blue, red, green and magenta symbols are known pulsators from
the literature, while black points mark \kic\ entries
(stars: WD candidates, crosses: sdB candidates, circles: other).
Note that while single sdB stars are exceedingly blue, sdB stars with
F/G/K-star companions overlap with the general population of blue stars.}
\label{fig:rpm}
\end{figure}

\subsection{The SDSS/{\sevensize\bf SEGUE} sample}
The {\sc segue} extension \citep{SEGUE} of the Sloan Digital Sky Survey \citep[SDSS,][]{SDSS}
contains a single stripe that crosses the south-western part of the
\kep\ target field. Searching the Sloan database for UV-excess targets revealed 19 acceptable
targets brighter than $g$\,=\,17.5. Of these, three have spectroscopy in SDSS/{\sc segue} identifying
them as hot subdwarfs. But selecting on the near-optical $u$ color is as effective in
distinguishing compact stars from normal stars as using {\em Galex} far-UV colors; all
ten stars selected from the current sample with this method turned out to be hot subdwarfs
or white dwarfs.

\subsection{The reduced proper motion sample}
Nineteen candidates were selected using the reduced proper motion (RPM) method, which is a
powerful tool to identify white dwarfs in photometric surveys.
RPM is defined as (\mkep\,+\,5\,log\,$\mu$\,+\,5), where $\mu$ is the proper motion in arcsec/yr
from the USNO catalogue. When plotted against the $g-i$ color from the \kic, hot
high proper motion objects are easily isolated (see \figref{fig:rpm}).
However, the $g-i$ colour is not a precise
temperature indicator and therefore it is difficult to say whether these
targets fall inside one of the known instability strips.
On the other hand, unknown instability strips might exist as shown by recent
discoveries: the first (and up to now unique) sdO pulsator \citep{woudt06}
and the first hot DQ oscillating white dwarf \citep{montgomery08}.

For hot subdwarfs the method has a lower likelihood of producing
successful results, since these stars are typically much more distant
than white dwarfs,
and the USNO proper motions are not precise enough to be useful in most
cases. Five of the nineteen candidates are included in the current survey sample, and our
spectroscopy has confirmed that four of them are indeed white dwarfs, and the remaining
one is an sdB star.

\subsection{Targets selected on {\sevensize\bf KIC} colours}
The \kep\ Input Catalog (\kic)~contains $g-r$ colours that can be used to select blue stars.
However, without a UV color it is not really possible to distinguish between normal B stars
and hot subdwarfs, in particular because one third of sdBs are found in binary systems
with F\,--\,K companions, and are therefore redder than they would be as single stars.
However, since B stars are much more luminous than subdwarfs and white dwarfs, below a magnitude
around $\sim$15, the limited thickness of the galactic disc ensures that any blue star is much more
likely to be a compact object than a main-sequence or regular horizontal branch star.

Two of the proposals used this strategy for the bulk of their targets, and
unfortunately only one of those had access to the results of the \twom\ survey. Thus, eight stars
known to be normal B stars entered the current sample. However, of the remaining thirteen stars 
selected with this method, ten turned out to be compact objects from our spectroscopy.

\begin{table}
\caption[]{Log of spectroscopic observations.}
\label{tbl:spectro}
\centering
\begin{tabular}{llrl} \hline
Date                  & Tel-run&$N_t$ & P.I.~+ Observer(s) \\ \hline
2008 June--September  & Bok    & 15 & EMG \\
2008 August 16        & INT-1  &  2 & CRL, MV \\
2008 September 20--24 & NOT-1  & 19 & RO \\
2009 April 11--12     & WHT-1  & 37 & CA, RH{\O} \\
2009 June 5--10       & INT-2  &  2 & RO, RH{\O}, TAO\\
2009 July 14--16      & WHT-2  & 32 & CA, TAO \\
2009 August 14        & NOT-2  &  2 & JHT, TL \\
2009 September 7      & NOT-3  &  5 & JHT \\ \hline
\end{tabular}
{\sc Note.}---$N_t$ lists the number of targets observed during the run
that actually entered the \kep\ compact star survey sample.
\end{table}

\section{Ground-based observations}\label{sect:spectro}

Throughout the period leading up to the release of the first \kep\
data we undertook preliminary ground-based photometry and spectroscopy
of most of the targets in the \kep\ sample, in order to improve
the classifications from the initial surveys.

\subsection{Spectroscopy}
We obtained classification spectra of all the targets in
Table~\ref{tbl:targets} using low resolution spectroscopy at
various telescopes, as listed in Table~\ref{tbl:spectro}.

The observations at the {\em Isaac Newton Telescope} (INT) used the
IDS spectrograph
with the 235\,mm camera and the R400B grating, providing a resolution
of R\,$\approx$\,1400 and an effective wavelength coverage
$\lambda$\,=\,3020\,--\,6650\,\AA.
At the {\em Nordic Optical Telescope} ({\sc not}) we used the
{\sc alfosc} spectrograph, with 
grism \#6 in 2008, and grism \#14 in 2009. Both give $R$\,$\approx$\,600
for the wide slit we used, and $\lambda$\,=\,3300\,--\,6200.
On the {\em William Herschel Telescope} (WHT)
we used the {\sc isis} spectrograph with grating R300B on
the blue arm (R\,$\approx$\,1600, $\lambda$\,=\,3100\,--\,5300\,\AA).
Red arm spectra were also obtained, but have not been used for this
work.
Observations from the {\em Bok telescope} on Kitt Peak were done with the B\&C
spectrograph using a 400/mm grating ($R$\,$\approx$\,550,
$\lambda$\,=\,3620\,--\,6900\,\AA).

All data were reduced with the standard {\sc iraf} procedures for
long-slit spectra. The bias levels were estimated from the overscan region
of each image, and a large number of flat-field images were averaged
in order to calibrate the pixel-to-pixel response. Arc reference spectra
were taken frequently during the observations, and each target spectrum was
wavelength calibrated with the closest available arc. For some of the
observation runs no radial velocity (RV) or flux standards were obtained,
as the purpose of the observations was to produce classification spectra
that do not require such detailed calibration.

\subsubsection{Spectral classification}

All stars were classified spectroscopically to distinguish
the compact pulsator candidates from contaminating objects.
In the sample of 63 stars, only twelve were found to have spectra
indicating that they are main-sequence or horizontal branch stars
(the classifications are given in Table~\ref{tbl:targets}).
It is worth noting that almost all of the main-sequence B stars that
contaminate our sample are variable, although with much
longer periods than would be exhibited by our compact objects.
However, we will not discuss these stars further in this paper.

\begin{figure}
\centering
\includegraphics[width=\hsize]{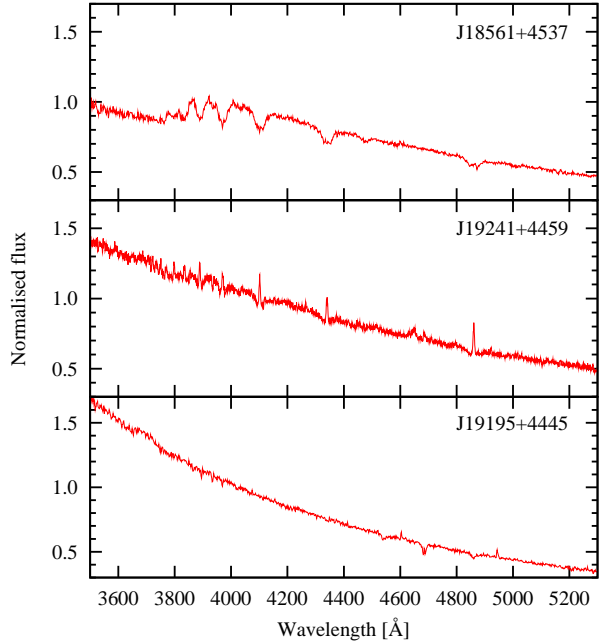}
\caption{Spectra of two cataclysmic variables and one planetary nebula nucleus (bottom).}
\label{fig:hot}
\end{figure}

Three unusual objects are worth particular mention, and we show their
spectra in \figref{fig:hot}.
The slopes of these spectra were rectified from the instrumental response
to a normalised flux scale by dividing the spectrum with a response function
computed by calibrating an extinction corrected spectrum of a single sdB star
against a flux model spectrum for its atmospheric parameters.

\begin{figure*}
\centering
\includegraphics[width=12cm]{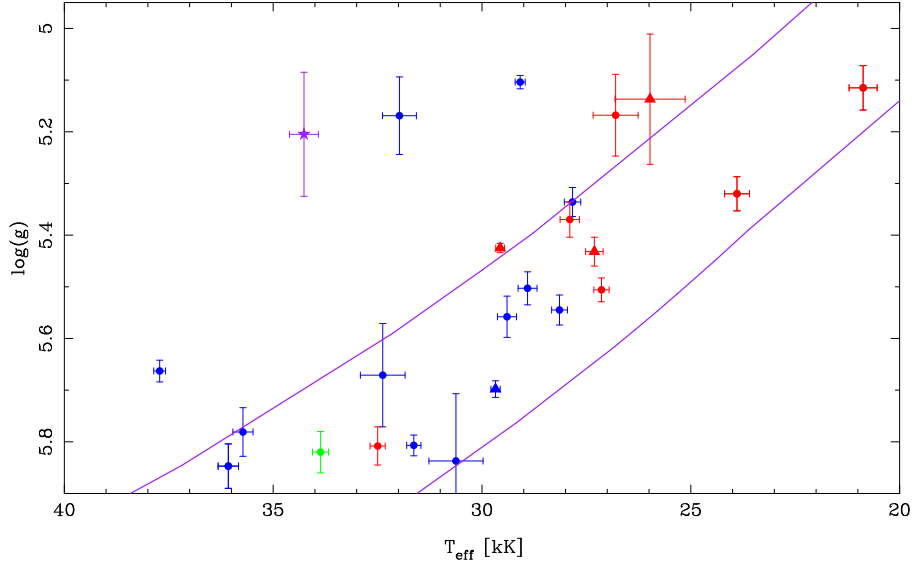}
\caption{The \teff/\logg\ plane for the sdB stars in our sample.
Red symbols indicate the pulsators, blue symbols the non-pulsators, with
the transient pulsator marked with a green symbol. The sdB+dM reflection
binaries are marked with triangles, and the eclipsing sdB+WD system
(which is also a candidate pulsator) is marked with a star. }
\label{fig:tgplot}
\end{figure*}

J18561+4537 shows Balmer line cores apparently filled in by emission,
and J19241+4459 shows Balmer lines in clear emission, and we classify both
objects as cataclysmic variable (CV) systems. 
They are both strongly variable in the \kep\
photometry, and we give a brief description of these objects
in Section~\ref{sect:cvs}.

J19195+4445 is an extraordinary
hot object. Very few features are visible in its hot continuum,
but the combination of C\,{\sc iv} and He\,{\sc ii} close
to 4686\,\AA\ is a hallmark of the PG1159 stars.
\citet{kronberger06} have noted a `possible bipolar PN' which they
call DSH\,J1919.5+4445, with an area
of about two arcmin centered roughly on the coordinates of our object.
We conclude that J19195+4445 is the central star of that nebula.

\subsubsection{Properties of the hot subdwarfs}\label{sect:sdbs}

We have determined the physical parameters of all the sdB stars in our sample,
except for two that are strongly contaminated by a companion
(listed as sdB+F/G in Table~\ref{tbl:targets}).
The physical parameters derived from our model fits are listed in the last columns of
Table~\ref{tbl:nonpuls} and Table~\ref{tbl:pulsators}.
Almost all the sdB stars show the \Caii\ H and K lines characteristic of sdBs with
main-sequence F/G/K companions, but in our low resolution spectra it is not
possible to distinguish the \Caii\ signature of a main-sequence companion from
interstellar absorption, which can be quite substantial at the relatively low
galactic latitudes of the \kep\ field. Only J19162+4749 and J19405+4827
are clearly composite-spectrum sdB plus F or G binaries, and
we refrain from attempting to determine physical parameters for these two
stars due to the strong contamination.
One star, J19415+4907, was classified as sdOB based on the presence of both
He\,{\sc i} and He\,{\sc ii} lines in its spectrum, but we failed to obtain
a satisfactory solution in the model-fitting procedure.
The object appears to be hotter than 40\,000\,K, but a
higher S/N spectrum would be required to make firm conclusions.

The effective temperature (\teff), surface gravity (\logg),
and photospheric helium abundance (\logy\,= \lheh) were derived by
fitting model atmosphere grids to the hydrogen and helium lines visible in the
spectra, which in all cases include Balmer lines up to H$_{13}$. 
The fitting procedure used here was the same
as that of \citet{edelmann03}, but we used only the metal-line blanketed
LTE models of solar composition described in \citet{heber00}.
Note that for the hottest stars in the sample, NLTE effects start to
become significant \citep{napiwotzki97}, and that the LTE models can
underestimate the temperatures by up to 1000\,K in some cases.
It is also known that sdB stars have very peculiar abundance patterns.
In general the light elements (He to S) are deficient with respect to the sun
\citep{edelmann06},
while some iron-group and even heavier elements ({\em e.g.} lead) can be
strongly enriched \citep{otoole06,blanchette08}.
The iron abundance, however, is found to be nearly solar
(with a scatter $\pm$0.5\,dex from star to star)
in a sample of 139 sdB stars, irrespective of the effective temperature
\citep{geier10a}. Therefore, we decided to use model atmospheres of
solar metal content, which account for metal-line blanketing through opacity
distribution functions, for the quantitative spectral analysis
\citep{heber00}. The helium abundance is determined for each star
individually.
The peculiar metal abundance patterns are, hence, unaccounted for and might
introduce some uncertainties when determining \teff\ and \logg.
The errors given on the last digit in Table~\ref{tbl:nonpuls} and
\ref{tbl:pulsators} only reflect the formal errors when fitting the
models to the observed spectra. Systematics when using grids with different
numerical approximations or metallicities can be much more significant
than the formal errors, especially for the stars with high S/N spectra,
which appear with extremely low error bars in \figref{fig:tgplot}.

The parameters of the pulsators are given in Table~\ref{tbl:pulsators},
together with their variability data. Remarkably, all the stars in our sample
which converge to temperatures below $\sim$27\,500\,K on our LTE grid
appear to be pulsators (see \figref{fig:tgplot}). 
Several stars lie in the region close to 28\,000\,K where the hybrid
DW\,Lyn type pulsators are found.
Three of the long-period pulsators also
show signs of pulsation periods with frequencies short enough to
be $p$-modes, although with extremely low amplitudes.

\begin{table*}
\caption[]{Properties of sdB stars with no significant pulsations.}
\label{tbl:nonpuls}\centering
\begin{tabular}{rlrrrrrrrrrcccl} \hline
\multicolumn{2}{c}{}    &
\multicolumn{3}{c}{100\,--\,500\,\uHz} &
\multicolumn{3}{c}{500\,--\,2000\,\uHz} &
\multicolumn{3}{c}{2000\,--\,8488\,\uHz} &
\multicolumn{4}{c}{Spectroscopic data} \\
         &              &\tabsig&\Amax&\Fmax &\tabsig &\Amax&\Fmax
                        &\tabsig &\Amax&\Fmax &\teff &\logg  &\logy   & Tel-run\\
\tabkic  & Name         &ppm&\tabsig&\uHz&ppm&\tabsig&\uHz&ppm&\tabsig&\uHz&kK&dex&dex&\\ \hline
 9569458 & J18477+4616  & 92&4.0& 195& 91&3.5& 881& 92&3.7&5516&27.8(2)&5.34(3)&--2.7(1)& WHT-2 \\
 4244427 & J19032+3923  & 97&3.5& 245& 95&4.1&1253& 95&4.0&2903&28.1(2)&5.55(3)&--2.0(1)& WHT-2 \\
 8022110 & J19179+4350  & 42&2.9& 146& 41&3.5&1322& 41&3.5&6324&28.9(2)&5.50(3)&--2.6(1)& WHT-2 \\
 6848529 & \bd          &  3&2.8& 473&  2&3.3&1880&  2&3.8&4887&28.7(1)&5.08(2)&--1.6(1)& WHT-1 \\
         &              &   &   &    &   &   &    &   &   &    &29.5(1)&5.13(2)&--1.5(1)& Bok \\
10661778 & J19211+4759  & 99&3.5& 452& 97&3.5& 805& 98&4.1&2914&29.4(2)&5.56(4)&--2.8(1)& WHT-2 \\
 1868650 & KBS\,13      &  7&3.4& 297&  7&3.6&1365&  7&3.7&5033&29.7(1)&5.70(1)&--1.7(1)& Bok \\
 3729024 & J19014+3852  &137&3.1& 200&133&3.5& 693&134&3.8&2587&30.6(7)&5.83(9)&--3.2(1)& SDSS \\
 6188286 & J19028+4134  & 10&4.0& 188& 10&4.1&1206& 10&3.9&4921&31.6(2)&5.81(2)&--1.7(1)& WHT-1 \\
 7434250 & J19135+4302  & 20&4.0& 183& 19&3.4& 576& 19&3.9&6196&32.0(4)&5.17(7)&--2.8(2)& WHT-1 \\
 8889318 & J19328+4510  & 62&3.4& 412& 61&3.7&1614& 62&3.8&3843&32.4(5)&5.67(9)&--2.4(4)& WHT-1 \\
 8142623 & J18427+4404  &118&3.0& 274&115&3.7& 653&116&4.0&4291&34.1(2)&5.65(3)&--1.5(1)& WHT-2 \\
 7975824 & KPD\,1946+4340& 9&3.2& 275&  9&3.7&1156&  9&4.1&5018&34.3(3)&5.20(12)&--1.3(1) \\
 5342213 & J18568+4032  &157&3.6& 182&155&3.3&1620&156&3.6&4033&35.7(2)&5.78(5)&--1.5(1)& WHT-2 \\ 
 8496196 & J19288+4430  & 41&3.1& 323& 41&3.7&1189& 40&3.9&8065&36.1(2)&5.85(4)&--1.6(1)& WHT-2 \\
 9543660 & 2M1951+4607  &  7&3.4& 221&  7&3.6&1994&  7&4.0&3826&37.7(1)&5.66(2)&--2.3(1)& Bok \\
10593239 & J19162+4749  &\multicolumn{3}{c}{Var.~comp.}
                                     & 18&3.3& 576& 18&3.7&5732&\multicolumn{3}{c}{Strong composite}& NOT-1 \\
10982905 & J19405+4827  &  9&3.3& 314&  8&3.5&1098&  8&3.7&4797&\multicolumn{3}{c}{Strong composite}& NOT-1\\
 \hline
\end{tabular}
\end{table*} 

\subsection{Photometric observations}\label{sect:photobs}
Prior to the \kep\ launch, ground-based preliminary observations were done for 27 targets,
including all those with known proper motion.
Time-series photometry on 24 of these was performed at the 3.6-m
{\em Telescopio Nazionale Galileo} (TNG) in August of 2008 using the
{\sc dolores} camera (marked with a $\dag$ in Table~\ref{tbl:targets}).
Each target was observed once for one to two hours with the $g$ filter,
and exposure times between one and ten seconds, depending on the magnitude.
The data were reduced using standard techniques \citep[see e.g.][]{silvotti06},
using typically seven comparison stars (minimum four).
A preliminary analysis of these data did not show any clear signature of
intrinsic
pulsation in any of the stars observed \citep{silvotti09} and the typical upper
limits to the pulsation amplitude were between 0.1 and 0.2\%.
However, for two sdB candidates low-amplitude variability was suspected:
a peak at 125\,s with an amplitude of about 0.1\% at 3.5 times the noise level
was found in \kic\,2020175 (=\,J19308+3728, a Q3.1 \kep\ target)
and a possible period of about one hour with an amplitude of $\sim$0.4\% was
detected in KPD\,1943+4058.
Indeed, \kep\ has confirmed that KPD\,1943+4058 is a $g$-mode pulsator
(see Table~\ref{tbl:pulsators} and Section~\ref{sect:puls} below).
Moreover, the TNG data revealed high amplitude ($\sim$10\%) variability for
the CV J19241+4459 (see Section~\ref{sect:cvs} below),
and ground-based follow-up photometry and spectroscopy have recently been
published by \citet{williams10}.

Another three targets were observed with {\sc not}/{\sc alfosc} in fast photometry mode
(marked with a $\ddag$ in Table~\ref{tbl:targets}).
Two showed no particular features, and the photometric limits were
included in the study of \citet{sdbnot}, and one
star observed during the NOT-1 run
was found to show substantial photometric activity on longer time-scales,
but the one-hour run was too short to make any further conclusions.
The \kep\ photometry clearly reveals this object, J18561+4537 (\kic\,9202990) to
be a cataclysmic variable (see Section \ref{sect:cvs}, below).

\section{{\em Kepler} observations}

Both sdBs and WDs are rare objects in the field. The large PG and KPD surveys detected only
about five sdBs with $B<16$ per one hundred square degrees of sky, and half of
that for the WDs.
The FoV of \kep\ covers about that area, so in order to have a reasonable chance of
finding a pulsator, it was necessary to extend the limit to a magnitude of $\sim$17.5.
The high photometric accuracy of \kep\ ensures that we can still detect pulsators
at this magnitude, and therefore that compact pulsators remain viable targets for the mission.
Although the 1-min cadence is not ideal for the short periods often found in these
targets, it is short enough to detect most if not all the pulsation periods of these stars,
which are typically between about two minutes and two hours.

\subsection{Sample statistics}

Our initial statistics predicted about twelve sdBV
(four short-period and eight long-period sdBVs)
and two to three ZZ\,Ceti pulsators within the 17.5 magnitude limit.
The DBV and GW\,Vir stars are more rare, and would only be found by a
stroke of luck.
These numbers were obtained assuming a local space density of
$1.2\times10^{-4}$~pc$^{-3}$ (ZZ\,Ceti), $4\times10^{-8}$~pc$^{-3}$ (V361\,Hya) and
$8\times10^{-8}$~pc$^{-3}$ (V1093\,Her), a disc scale height of 500\,pc,
and considering a fraction of pulsators within each instability strip
of 1, 0.1 and 0.5 respectively \citep[see][for more details]{silvotti04}.
Note that, due to their intrinsic faintness, all the DAVs observable with \kep\
belong to the galactic disc, implying that their number is highly dependent on
the magnitude limit (while this is only marginally true for the sdBs).
The number of DAVs potentially observable with \kep\ increases by a factor
$\sim$4 when increasing the magnitude limit by one unit.

The initial statistics have now been confirmed to a large extent by our spectroscopic
observations, as our efforts have securely identified 70 hot subdwarfs down
to a magnitude of 17.7. As many as 56 of these are sdBs and sdOBs (exactly the
number expected from the space densities and pulsator fractions stated above),
seven are He-sdOBs, and seven are sdOs,
although the exact number of pulsators remains
to be determined as only half of the photometric survey data
have been released.
For the WDs the numbers are about as expected;
fifteen DAs, two DBs, one DO, two PNNi and two CVs.
About half of these targets are listed in Table~\ref{tbl:targets},
the remainder will be described following the release of the second
half of the survey phase of the mission.

\begin{table*}
\caption[]{Properties of the sdBV stars.}
\label{tbl:pulsators}
\centering
\begin{tabular}{rlccrrrrcclcl} \hline
\multicolumn{2}{c}{}    & 
\multicolumn{5}{c}{\kep\ data} & \multicolumn{4}{c}{Spectroscopic data} \\
\multicolumn{2}{c}{}    &$N_f$&\fmed &\fmin &\fmax &\porb&
\teff &\multicolumn{1}{c}{\logg}&\logy & Tel-run\\
\tabkic & Name         &     &\uHz  &\uHz  &\uHz  & days&
 kK   &\multicolumn{1}{c}{dex}& dex & \\ \hline
10670103& J19346+4758  &  28 & 131.6&  61.4& 203.2& ... &20.9(3)&5.11(4) &--2.2(3)&NOT-3\\ 
 2697388& J19091+3756  &  37 & 183.1&  89.1&3805.9& ... &23.9(3)&5.32(3) &--2.9(1)&NOT-1\\ 
11179657& J19023+4850  &  11 & 271.5& 186.5& 351.6&0.394&26.0(8)&5.14(13)&--2.1(2)&WHT-1\\
 7664467& J18561+4319  &   6 & 207.5& 110.2& 246.9& ... &26.8(5)&5.17(8) &--2.8(2)&WHT-2\\ 
 5807616&KPD\,1943+4058&  21 & 224.1& 109.5&3447.2& ... &27.1(2)&5.51(2) &--2.9(1)&Bok \\  
 2991403& J19272+3808  &   7 & 283.7& 194.7& 334.8&0.443&27.3(2)&5.43(3) &--2.6(1)&WHT-2\\
 3527751& J19036+3836  &  41 & 275.1&  92.1&3703.3& ... &27.6(2)&5.28(3) &--2.9(1)&NOT-1\\ 
        &              &     &      &      &      &     &28.2(3)&5.46(5) &--2.9(1)&INT-1\\
        &              &     &      &      &      &     &27.9(2)&5.37(9) &--2.9(1)&(Adopted) \\
 9472174 & 2M1938+4603 &$>$55&2065.9&  50.3&4531.6&0.126&29.6(1)&5.42(1) &--2.4(1)&Bok \\
10139564& J19249+4707  &  20 &5709.4& 316.0&7633.6& ... &32.5(2)&5.81(4) &--2.3(1)&WHT-2\\ 
 2991276 & J19271+3810 &   1 &8201.1&      &      & ... &33.9(2)&5.82(4)&--3.1(1)& WHT-2 \\
\hline
\end{tabular}
\end{table*}

\subsection{The {\em Kepler} photometry}\label{sect:readout}

\kep\ short-cadence (SC) photometry consists of light curves with 58.8\,s
sampling, each around one month in length, except for the shorter Q0
commissioning cycle which only included relatively bright objects. 
The light curves from Q0 and Q1 are virtually
uninterrupted, but in Q2 events on July 2 triggered a
processor reset that caused the spacecraft to enter safe mode.
Such safing events produce gaps in the light-curves, and the July 2
event caused a two-day gap in time-series from Q2.1. Two other minor events
caused a loss of fine pointing control, which produced a half-day gap in the
photometry from from Q2.2, and a one-day gap in Q2.3.
Pointing corrections that did not leave significant gaps also produced minor
excursions in the light curves. 
Such corrections have minimal effect on uncontaminated light curves,
but can cause quite significant excursions in the light curves of
stars with substantial contamination factors (see Table~\ref{tbl:targets}).
We fitted first or second order polynomials with an exponential decay
component to each segment of the light curve and subtracted this function
before restoring it back to its mean value. The amplitudes provided in
this paper are therefore not corrected for the contamination from close
companions. We are concerned that the contamination corrections provided
by the \kic\ are not sufficiently accurate for our targets, as
they have spectra that deviate substantially from normal stars, particularly
at short wavelengths.  A full contamination correction will
be performed later for the most important targets, when we have spectroscopic
information also on the surrounding stars.

The \kep\ instrument has an intrinsic exposure cycle consisting of 6.02\,s
integration followed by a 0.52\,s readout period. The SC photometry is a sum of
nine such integrations, and long-cadence (LC) photometry are a sum of 270
\citep{gilliland10b}.
For unknown reasons, the LC cycle produces artefacts in the SC light curve, not
just at the LC frequency, \flc\,=\,566.391\,\uHz, but at all harmonics
of this frequency up to the Nyquist frequency, which is
\fnyq\,=\,15$\cdot$\flc\,=\,0.5$\cdot$\fsc\,=\,8496.356\,\uHz.
In most short-cadence
light curves, these artefacts get stronger at higher frequencies,
typically peaking at 9$\cdot$\flc\,=\,5098\,\uHz.
After the safing event in Q2.1, \flc\ returned with a different phase, which
makes it split up in two peaks of roughly equal amplitude when taking
the FT of the full dataset, while stellar pulsation peaks remain single-peaked.
In these cases it affects a region of the FT up to 2\,\uHz\ wide
around each multiple of \flc,
while for all other runs the regions affected are $\sim$1\,\uHz\ wide.
Another likely artefact is found at 7865\,\uHz.
It is seen in the light curves of \kic\,8022110, and 2991276, in the pulsators
\kic\,2697388, 3527751, 10139564, and in the DB white dwarf \kic\,6862653.
\kic\,2991276 was observed in Q2.1, and the 7865-\uHz\ peak is double
just as the \flc\ artefacts.
It thus appears most likely that this peak is instrumental in origin,
but its particular cause is not yet known.

\begin{figure*}
\centering
\includegraphics[width=\hsize]{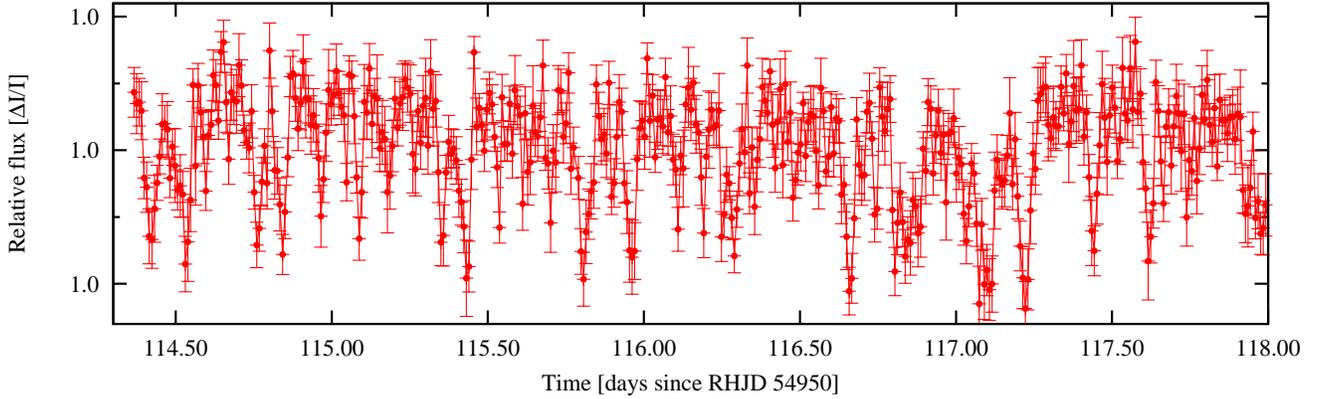}
\caption{The light-curve of the He-sdOB J19352+4555 shows unusual dips which appears as
1/$f$ noise in the FT, \ie\ there is no regular recurrence. Here, only the first 3.7 days
of the 1-month dataset is shown. Since there is no short-term variability, the data have
been binned with each point representing the average of twelve short-cadence exposures, and
their rms variation.}
\label{fig:J19352_lc}
\end{figure*}

\subsection{Detections and non-detections}\label{sect:sdbphot}

Unambiguous pulsations were detected in nine of the stars classified as hot
subdwarf stars in Table~\ref{tbl:targets}, and in none of the white dwarfs.
For this reason we will concentrate mostly on the subdwarf\,B stars in this paper,
and leave a detailed treatment of the white dwarfs until the survey is complete.
After detrending the light curves,
we computed their Fourier transforms (FTs) and examined them 
for pulsations and binary signatures. The pulsators are discussed in
Section~\ref{sect:puls}, and the binaries in Section~\ref{sect:bins}, below.

Table~\ref{tbl:nonpuls} and Table~\ref{tbl:nonsdb} list the limits from the
\kep\ photometry, in three different frequency ranges,
for all the stars where no clear pulsations were found.
We give the arithmetic mean of the amplitude spectrum
in the frequency range considered (which for a photon noise
dominated signal can be considered as the variance, \sig),
and the amplitude, \Amax, and frequency, \Fmax, of the highest peak, where
\Amax\ is given as the ratio of the peak amplitude and the \sig\ level.

With only a few exceptions, the frequency analysis of the three frequency
regions yields identical results in terms of noise level,
evidence of the excellent quality of the \kep\ photometry.
The noise level is found to be $\sim$8\,ppm for \mkep\,=\,14 and
$\sim$100\,ppm for \mkep\,=\,17.5. Note that there are some discrepancies due
to the varying contamination, which is not reflected in the \kep\
magnitudes.  The highest peak found in the FT for the targets in
Table~\ref{tbl:nonpuls} lies on average
around 3.7 times this level, but there are quite a few stars that show
peaks between 4.0 and 4.1 times the mean. This is not unexpected considering
the fact that the light curves contain around 40\,000 independent measurements,
and it is hard to claim that such low-amplitude pulsations are significant.
However, it is worth noting that there seem to be more cases of such 4.0
times mean peaks in the sdB stars (Table~\ref{tbl:nonpuls}) than in the
non-sdBs (Table~\ref{tbl:nonsdb}). Since the sdBs are known to show pulsations
more frequently than the other classes, this may not be a coincidence, and
rather an indication of low-amplitude or transient pulsation modes.

For two stars we were unable to derive useful limits in the low-frequency
range.
The sdB+F composite J19162+4749 shows a light curve that is not consistent
with orbital motion (\figref{fig:J19162}). The main variation is found
at 1.9\,d, far too long for a compact pulsator. We will discuss this object
further in Section~\ref{sect:bins}.

The second object with substantial power in the FT at low frequencies is
the He-sdOB star J19352+4555. The light curve appears to have rapid drops
in the photometry with no apparent regularity (Fig.~\ref{fig:J19352_lc}),
and is unlike anything we have seen in any other \kep\ light curves.
It is listed with a contamination of 80 per cent, so the variation may not be
from the subdwarf, but we have no explanation for the cause.

For KBS\,13 and the other low-amplitude reflection binaries we were
able to subtract the
orbital effect with a simple two component sine function, to produce the limits
given in Table~\ref{tbl:nonpuls}.
For KPD\,1496+4340 we subtracted the optimal model from \citet{bloemen10}
to produce the tabulated limits, and note that the 4.1-\sig\ peak at
5018.2\,\uHz\ is just above our significance limit.

\begin{table*}
\caption[]{Photometric limits on the non-sdB stars.}
\label{tbl:nonsdb}\centering
\begin{tabular}{rlrrrrrrrrrl} \hline
\multicolumn{2}{c}{} &
\multicolumn{3}{c}{100\,--\,500\,\uHz} &
\multicolumn{3}{c}{500\,--\,2000\,\uHz} &
\multicolumn{3}{c}{2000\,--\,8488\,\uHz} & \\
 & & \tabsig&\Amax&\Fmax&\tabsig&\Amax&\Fmax&\tabsig&\Amax&\Fmax   & Class \\
\tabkic  & Name         & ppm &\sig & \uHz & ppm &\sig & \uHz & ppm &\sig & \uHz & \\ \hline
 2297488 & J19208+3741  &  78 & 3.5 & 186.8&  76 & 3.4 &1714.8&  77 & 3.8 &5174.3& sdO+F/G \\
 2692915 & J19033+3755  & 117 & 3.2 & 260.5& 112 & 3.5 &1492.1& 111 & 3.7 &8249.5& DO \\
 3427482 & J19053+3831  &  92 & 3.2 & 465.8&  93 & 3.5 &1430.4&  94 & 3.6 &3874.2& DA \\
 3527617 & J19034+3841  & 117 & 3.1 & 389.9& 116 & 3.5 &1383.7& 117 & 3.7 &4951.6& He-sdOB \\
 4829241 & J19194+3958  &  22 & 3.2 & 421.9&  22 & 3.9 &1900.9&  22 & 3.9 &8302.4& DA \\
 6669882 & J18557+4207  & 163 & 3.1 & 172.9& 161 & 3.4 &1316.1& 160 & 3.6 &6392.6& DA \\
 6862653 & J19267+4219  &  87 & 3.7 & 377.9&  83 & 3.3 &1143.1&  84 & 3.8 &7445.0& DB \\
 7353409 & 2M1915+4256  &  11 & 3.4 & 196.3&  11 & 3.5 & 884.0&  11 & 3.8 &5523.3& sdO \\
 7755741 & 2M1931+4324  &   8 & 3.1 & 176.5&   6 & 4.0 &1126.3&   7 & 3.8 &5387.3& sdO \\
 8619526 & J19195+4445  &  22 & 3.3 & 328.0&  22 & 3.8 &1708.1&  22 & 3.8 &5812.1& PNN \\
 8682822 & J19173+4452  &  23 & 3.7 & 344.0&  22 & 3.6 &1346.4&  22 & 3.6 &6031.4& DA \\
 9139775 & J18577+4532  & 147 & 2.9 & 186.2& 143 & 3.5 & 724.0& 144 & 3.7 &3490.4& DA \\
 9408967 & J19352+4555  &\multicolumn{3}{c}{High 1/$f$ noise}
                                           &  65 & 3.5 &1202.3&  66 & 3.5 &4954.2& He-sdOB \\
 9583158 & NSV\,11917   &  86 & 3.5 & 498.8&  81 & 3.5 &1303.3&  81 & 3.9 &6672.5& PNN \\
 9822180 & J19100+4640  &  12 & 3.3 & 282.8&  11 & 3.6 & 659.0&  11 & 4.7 &7444.2& sdO+F/G \\
 9957741 & J19380+4649  &  29 & 3.2 & 435.5&  29 & 4.0 &1162.9&  28 & 3.6 &4847.2& He-sdOB \\
10420021 & J19492+4734  &  25 & 4.5 & 196.4&  25 & 3.4 & 902.6&  25 & 3.9 &3748.9& DA \\
10482860 & J19458+4739  & 146 & 3.0 & 123.5& 145 & 3.7 & 686.1& 145 & 3.7 &5231.3& DO \\
11357853 & J19415+4907  &  75 & 3.4 & 222.9&  75 & 3.4 &1116.7&  76 & 4.1 &7009.0& sdOB (Hot)\\
11514682 & J19412+4925  &  25 & 3.4 & 213.3&  25 & 3.5 & 942.2&  25 & 3.5 &2822.0& DAB \\
11822535 & WD\,1942+499 &  15 & 3.0 & 208.4&  15 & 3.4 & 884.7&  15 & 3.5 &4991.4& DA \\
\hline \end{tabular}\end{table*}

\begin{figure}
\centering
\includegraphics[width=\hsize]{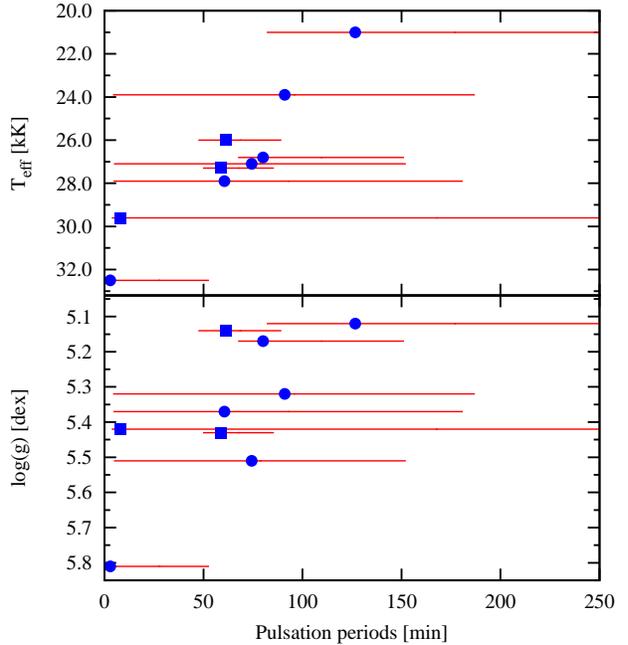}
\caption{Pulsation periods as a function of \teff\ and \logg\ for the
sdB pulsators from Table~\ref{tbl:pulsators}.
The red bars indicate the range of pulsations, and
the blue symbols mark the power-weighted mean frequency, \fmed.
The three short-period binaries are marked with boxes, the rest with bullets.
}
\label{fig:tgp}
\end{figure}

\subsection{Pulsation properties}\label{sect:puls}

One pulsator, \kic\,10139564 (J19249+4707),
stands out as the only unambiguous V361\,Hya pulsator
in the sample, and is the subject of a dedicated paper, Paper~{\sc ii}.
While having a pulsation spectrum with substantial power in the high frequency region,
a single peak with a frequency of 315.96\,\uHz~also appears in the FT.
This is surprising, as \kic\,10139564 is significantly hotter than the boundary
region where the DW\,Lyn stars are found, and much hotter than the theoretical
blue edge of the $g$-mode instability region.

The eclipsing binary 2M1938+4603 (\kic\,9472174) is an especially
interesting and unusual case. The pulsation spectrum is the richest of all
the sdBVs and spans the whole range from 50 to 4500\,\uHz, something never
seen before.  The only other eclipsing sdB+dM system with a
pulsating primary is NY\,Vir, so it is really an exceptional stroke of luck
that the brightest EHB star among all the compact pulsator candidates is
such an exceptional object. A first look at the pulsation spectrum and a
detailed analysis of the binary properties are given by \citep{2m1938}.

The remaining seven are all multi-periodic pulsators of the V1093\,Her class,
and a detailed frequency analysis will be provided in two separate papers.
Five objects: \kic\,2697388, \kic\,3527751, \kic\,7664467, \kic\,10670103, and
KPD\,1943+4058 are analysed in detail in Paper~{\sc iii}.
Interestingly, three of these show significant peaks in the short-period
pulsation region, making them likely hybrid $p$ \&\ $g$-mode pulsators as well.
For \kic\,3527751 and \kic\,5807616 this is not extremely surprising, as they
are close to the hot end of the $g$-mode region, but for \kic\,2697388 the
presence of a short-period pulsation is unexpected, as this is the second
coolest sdB star in the sample.
The first asteroseismic result based on one of the stars in Paper~{\sc iii}
is presented in Paper~{\sc iv}, and also
represents the first successful modeling
of a V1093\,Her star, leading through $g$-mode seismology to the
determination of the structural and core parameters of KPD\,1943+4058.

The two V1093\,Her stars, \kic\,2991403 and \kic 11179657, which 
clearly show a binary period in addition to their V1093\,Her type pulsation spectra
are the subject of Paper~{\sc v}.

In Table~\ref{tbl:pulsators} we list some general details from the frequency
analysis of these pulsators, together with the physical parameters derived
from our spectroscopy.
In addition to the number of significant pulsation modes, the minimum and
maximum pulsation frequency and the orbital period, we also provide the
power-weighted mean frequency, \fmed. The frequency range and \fmed\ are
plotted as a function of effective
temperature and gravity in \figref{fig:tgp}. 
Roughly linear, decreasing trends are evident in pulsation period as a function
of both \teff\ and \logg, but note that \teff\ and \logg\ are highly correlated
along the EHB (see \figref{fig:tgplot}).
These eight stars all show pulsations down to the significance limit in
their FTs, which means that
more frequencies are likely to be discovered in longer time-series.
All are scheduled for three consecutive months of short-cadence observations
in Q5, and will be followed closely over the full length of the \kepmi.
We defer the readers to the particular studies on these
objects for further details on their detailed asteroseismic properties.

\begin{figure}
\centering
\includegraphics[width=\hsize]{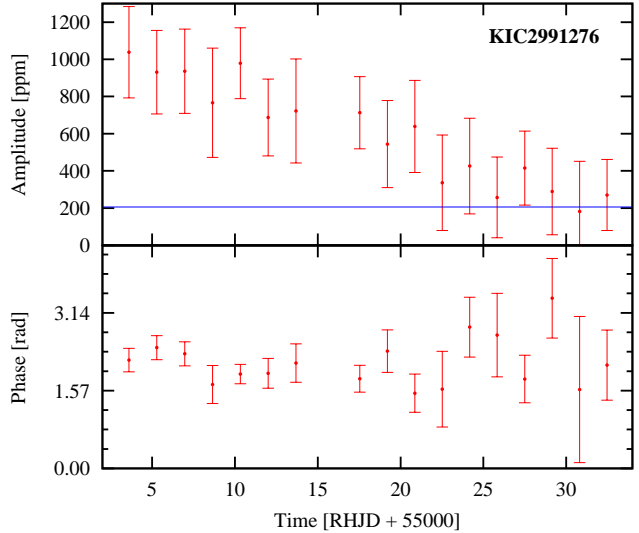}
\caption{The transient pulsation seen in \kic\,2991276.
The amplitude (upper panel)
and phase (lower panel) are shown for 17 roughly 1.7-day chunks of data.
The line in the upper panel indicates the average of the FT noise levels in the
chunks. }
\label{fig:trans}
\end{figure}

\subsection{Transient pulsations}

The FT of \kic\,2991276 shows a peak at 8201.1\,\uHz\ with an amplitude of
0.06\%, which is more than 10\sig.
At first we did not recognise this as a clear pulsator, as all other peaks
with similarly low amplitudes in the FT are associated with known artefacts.
However, no other stars show artefacts at this frequency, and since
\kic\,2991276
was observed during the Q2.1 run, all artefacts including the one of unknown
origin at 7865\,\uHz\ are double-peaked (see Section~\ref{sect:readout}),
and the 8201.1\,\uHz\ peak is not.

We examined the variability
of the amplitude and phase of this signal by dividing the one-month light curve
into 17 chunks and fitting the amplitude and phase of the frequency
determined from the complete data set. As \figref{fig:trans} shows,
the amplitude of the signal
decreased monotonically over the time-span of the observations, whereas
the phase remained constant within the errors. This argues against a
beating phenomenon as the phase would be expected to vary when the
amplitude approached zero.

At $P$\,=\,122\,s the true amplitude is substantially smeared during the
58.8\,s integration cycle. The smearing factor is given by $\sin (x)/x$, where
$x$\,=\,$\pi \Delta t_{\rm exp}/P$ \citep{kawaler93}.
In our case the FT recovers $\sim$66\%\ of the true amplitude.
But the contamination of this \mkep\,=\,17.4\,mag star from two stars located
eight arcseconds to the north (\mkep\,=\,13.8 and 14.2) is substantial. 
In the \kic\ \corr\ is given as 0.971, but as discussed above this is most
likely overestimated, since the \kic\ does not have the right flux distribution
for UV-excess stars.
Nevertheless, when the contamination and smearing are accounted for, the pulsation amplitude
in the beginning of the run must be several percent.


\kic\,2991276 is a normal He-poor sdB star with a temperature around 33\,900\,K
and \logg\,=\,5.8, which places it right in the middle of the instability
region.
Stars in this region of the instability strip often show variable
amplitudes and in rare occasions transient pulsation peaks \citep{kilkenny10a}.
A high amplitude mode at 122\,s is not unheard of, and quite comparable to the
main mode in QQ\,Vir \citep{silvotti02a}.
Single mode pulsators are also not unheard of in this temperature region.
LS\,Dra was observed to have a single pulsation mode with a period of
139\,s by \citet{ostensen01b}, and confirmed to be mono-periodic by
\citet{reed06}, but with a highly variable amplitude changing from a maximum
of 0.526\% to a minimum of 0.087\% over their campaign spanning 47 days.
Whether such amplitude variability is due to the beating of extremely closely
spaced modes, or due to true intrinsic amplitude variability requires
sustained observations over a much longer period. The \kepmi\ is
the ideal tool for accomplishing this.

\begin{figure}
\centering
\includegraphics[width=\hsize]{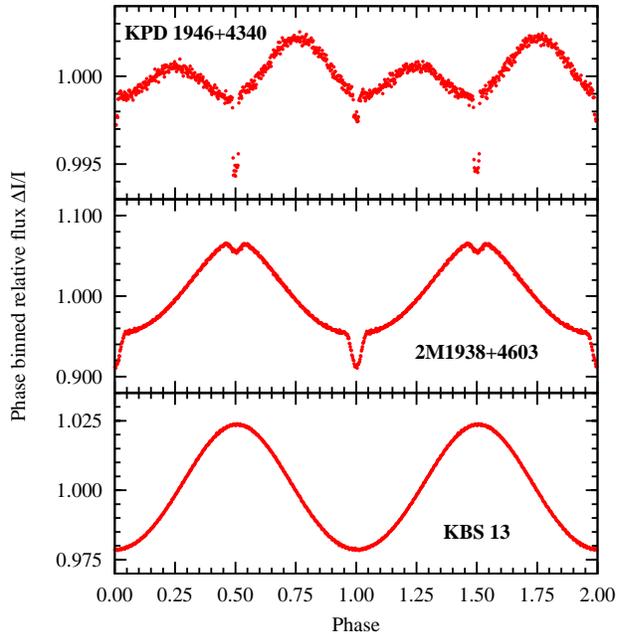}
\caption{Three binary sdB stars. The light curves have been folded on the
orbital period ($P$\,=\,0.404\,d, 0.126\,d and 0.29\,d respectively,
top to bottom), and averaged into 500 bins. Two cycles of the orbital
variation are shown.}
\label{fig:refl}
\end{figure}

\begin{figure}
\centering
\includegraphics[width=\hsize]{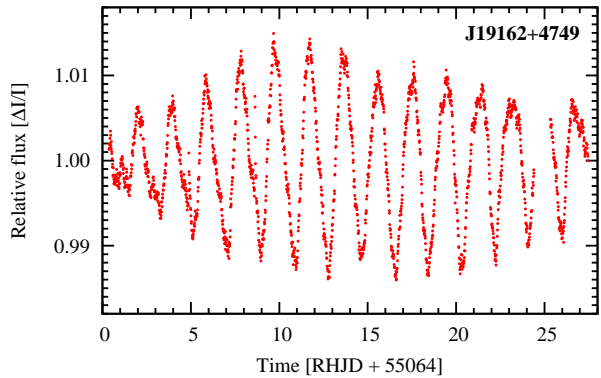}
\caption{The light curve of J19162+4749, a strong sdB+F composite.
Each point represents an average of 20 \kep\ short-cadence observations to
reduce the noise. The variability most likely comes from the F star.
}
\label{fig:J19162}
\end{figure}

\subsection{Binaries and long-period variables}\label{sect:bins}

\begin{table}
\caption{Binaries and other long-period variables.}
\label{tbl:bins}
\begin{tabular}{lllll} \hline
Name          &Period& Amp.& Class & Main \\
              & days & \%  &       & variability \\ \hline
KPD\,1946+4340& 0.404& 0.5 & sdB+WD & Eclipses \\
KBS\,13       & 0.29 & 2.5 & sdB+dM & Refl.~effect \\
J19135+4302   & 4.24 & 0.3 & sdB+? & Unknown \\
J18427+4404   & 1.12 & 1.5 & sdB+? & Unknown \\
J19405+4827   &$\sim$1& 0.1 & sdB+F & Unknown \\
\bd           & 1.09 & 0.03& sdB?  & Unknown \\
J19162+4749   & 1.9  & 0.9 & sdB+F & Companion \\
J19000+4640   &$\sim$2& 0.1 & sdO+F/G & Companion\\
J18561+4537   & 0.17 &  20 & DA+dM  & CV \\
J19241+4459   & 0.12 &  20 & DA+dM  & CV \\
J19193+5044   & 3.67 &  40 & DA+dMe & Flares \\
J19195+4445   & 1.12 & 0.05& PNN & Unknown \\ \hline
\end{tabular}
\end{table}

Five stars in our sample show clear long-period variability that we have
identified as binary in origin.
In addition to the three sdBs noted as pulsators in the previous section,
two non-pulsators are clearly identified.
Five further sdBs show variability that is unlikely to be intrinsic pulsations
in the hot subdwarf, and the sdO+F/G binary shows very low-level variations at
both short and long periods.
We have also noted
four variable systems with white dwarf primaries (see Table~\ref{tbl:bins}).
Note that the amplitude levels are only given as an indication of the significance
of the variability. They represent the peak Fourier amplitude of the raw light curves,
and are affected by contamination from nearby stars, which may also
be the source of the variability in some cases.

The most spectacular binary star in the sample is clearly KPD\,1946+4340.
This star was known to be an sdB+WD binary system from the RV study of \citet{morales-rueda03},
with a period of 0.4037\,d and an RV amplitude of 167\,km/s.
The \kep\ light curve folded on the orbital period (see top panel of \figref{fig:refl})
reveals that the object is clearly eclipsing, and shows a substantial ellipsoidal
deformation superimposed on a relativistic beaming effect. A detailed study of
this system is given by \citet{bloemen10}, which presents for the first
time a comparison between a radial velocity amplitude as determined from
spectroscopy with one determined from photometric relativistic effects.

The system 2M1938+4603 shows a light curve similar to the HW\,Vir
systems, except that the eclipses are much more shallow, indicating that
they are only grazing (\figref{fig:refl}, middle panel). 
This system was identified in 2008
during follow-up studies of hot subdwarfs from the \twom\ catalogue, and a
first analysis of the \kep\ data together with extensive ground-based
photometry and spectroscopy are provided in \citet{2m1938}.
The reflection binary KBS\,13 (bottom panel in \figref{fig:refl})
has also been known for some time already, and
details on the system were published by \citet{for08}.
With an RV amplitude of only 22.8\,km/s, the inclination angle
must be quite low unless the secondary is substellar.

\begin{figure}
\centering
\includegraphics[width=\hsize]{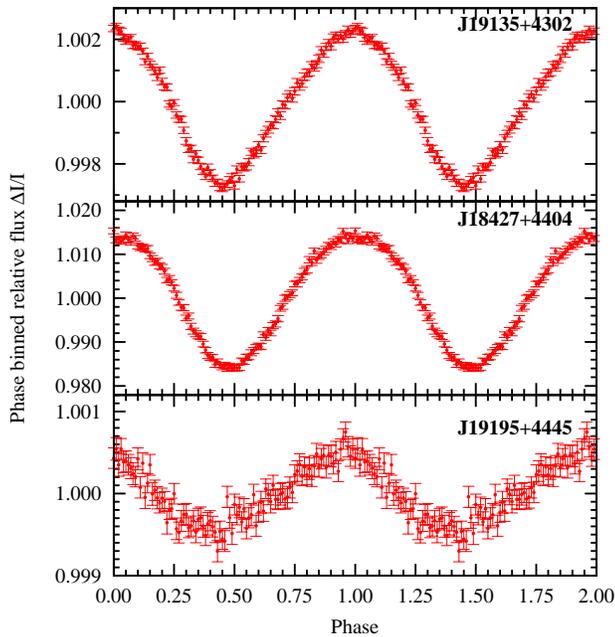}
\caption{Three objects showing low-level variability with regular
amplitude throughout the \kep\ run.
The light curves were folded on the main period
($P$\,=\,4.24\,d, 1.116\,d, and 1.119\,d respectively)
and binned into 100 bins to reduce the noise.
The variation is either intrinsic to a
companion or a contaminating field star, or
an orbital variation.
}
\label{fig:sdbins}
\end{figure}

The light curve of J19162+4749 (\figref{fig:J19162}) reveals substantial 
long-period variability,
but the amplitude of the $\sim$1.9-d period is clearly not constant.
The spectrum is composite with roughly half the optical contribution coming
from an F-star companion.  The target has a visible companion five
arcsec to the east,
but this star does not contribute more than $\sim$10 per cent of the light,
and cannot be the cause of the strong contamination
seen in the spectrum, so the system is most likely an unresolved binary with
an early F companion. The pulsations could be from the F star, in which case
it would be of the $\gamma$ Doradus class, but they
are sufficiently low-level that they could also be intrinsic to the contaminating
star. If they were from the F star, this would be the first sdB star found to
have a pulsating close companion. After concluding that there is no short-period
variability in this system, we have submitted the object for further long-cadence
\kep\ photometry. If $\gamma$ Doradus pulsations are confirmed in
the F companion of the sdB star, asteroseismology based on five years of
long-cadence \kep\ observations might shed light on the mass transfer phase
where the main-sequence star most likely received most of the envelope from
the sdB progenitor as it ascended the RGB.

Two more sdBs show slow amplitude modulation that might indicate
a binary nature, as listed in Table~\ref{tbl:bins}.
The clearest of these is J18427+4404 with an amplitude of
$\sim$\,1.5\% (middle panel of \figref{fig:sdbins}).
This star is not located in a very crowded field, but contamination
cannot be completely ruled out, so a spectroscopic series to determine its
radial velocity variation is required to firmly establish its binary nature.
There are
weak lines in the spectrum of this star that may be indication of an F/G/K
main-sequence companion, but this would be unusual as such binaries have
never been seen to have short orbits. The 1.12-d period has a constant
amplitude throughout the \kep\ run, and folding the light curve on this
period reveals that the light curve is not entirely symmetric, which
indicates either a slightly elliptic orbit, or some kind of low
level pulsation from a main-sequence contaminating star.
The same is the case for J19135+4302 (upper panel of \figref{fig:sdbins})
which shows an even more asymmetric cyclic variation, with a period of
4.24\,d.

\begin{figure}
\centering
\includegraphics[width=\hsize]{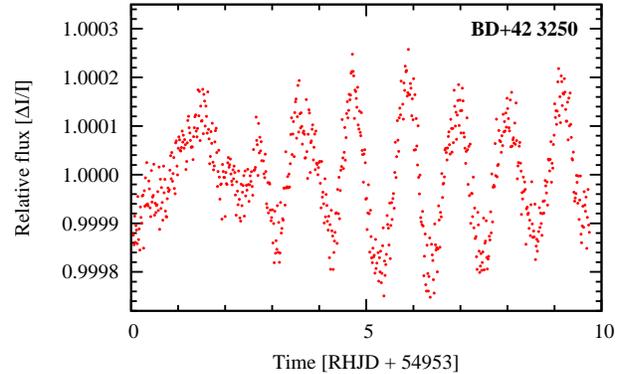}
\caption{The light curve of \bd\ shows extremely low amplitude variability
with periods of around one day.
20 \kep\ short-cadence observations were averaged for each point in the figure.
}
\label{fig:BD+42}
\end{figure}

The sdB star J19405+4827 shows irregular variations with many different
periods longer than half a day. This star is also a strong composite, so
the variability could come from the companion star, possibly from starspots.
Long-cadence data will be suitable to study the nature of these variations.

\begin{figure*}
\centering
\includegraphics[width=\hsize]{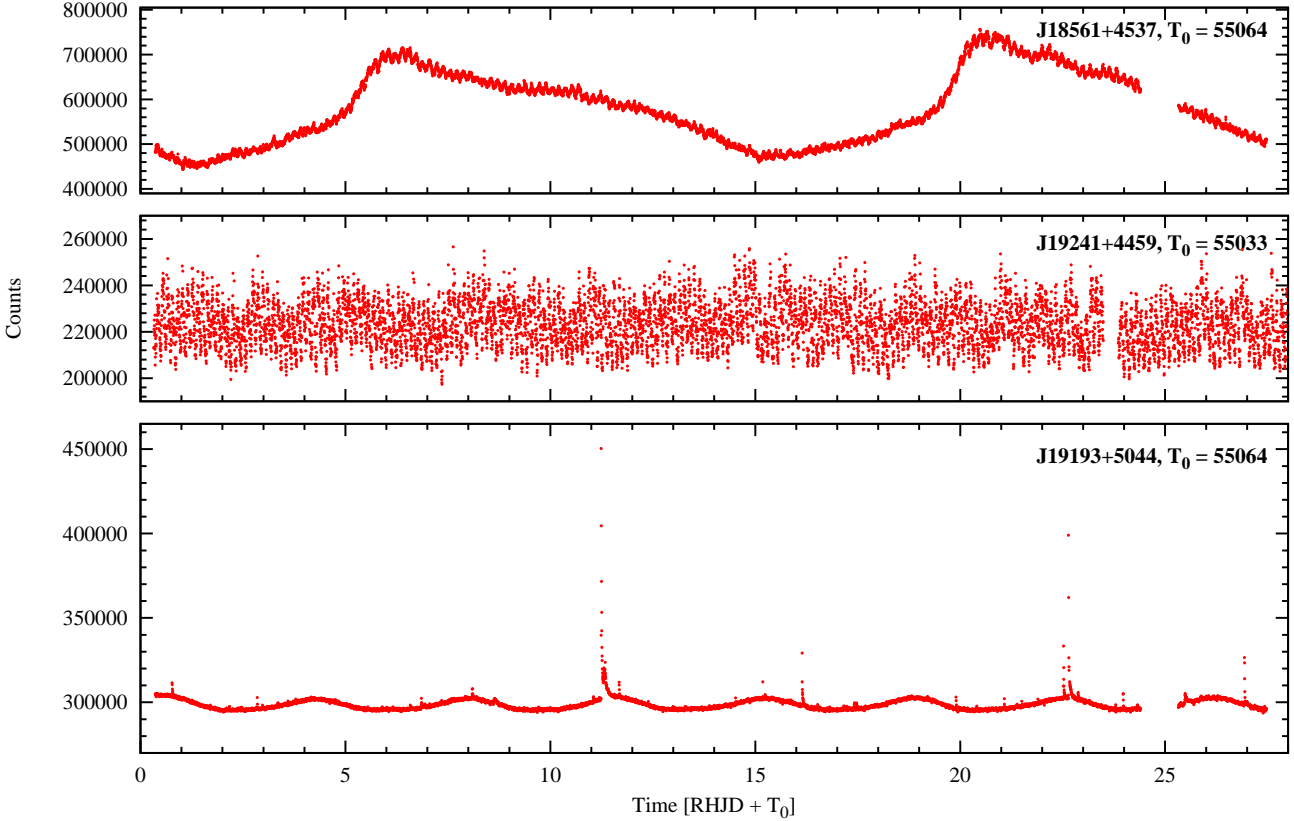}
\caption{High amplitude variable compact objects in binary systems.}
\label{fig:bins}
\end{figure*}

The same concerns about contamination apply to an even higher degree
for the lower amplitude systems with possible binary periods around one day.
The most unlikely to suffer contamination would be the bright
subdwarf \bd\ (\figref{fig:BD+42}),
as its contamination factor is only 0.003. However, the amplitude of 300\,ppm
is so low that a very substantial pulsation amplitude in one of the faint
neighbors can be invoked to explain the observed modulation. 
However, \bd\ is not a regular EHB star as evidenced by its low surface gravity
(\logg\,=\,5.1). This places it in a sparsely populated location in the
\teff,\,\logg\ plane, well above the canonical EHB. Post-EHB tracks for
shell-burning stars only reach
such low gravities when the temperature exceeds 32\,000\,K and post-RGB tracks
require a companion in order to bypass the horizontal branch stage.
Models of helium-core white dwarf mergers also pass through the hotter
parts of the \teff/\logg\ plane before reaching the EHB, but unlike
\bd, these merger products are expected to display enhanced atmospheric
helium at least until they settle on the EHB.
Since \bd\ is such an enigmatic object, the low-level variability
seen in the \kep\ photometry is worth further study.

The sdO+F/G binary J19100+4640 (\kic\,9822180) is a peculiar case, with
significant variations on long time-scales, and a single peak at 7444.2\,\uHz.
The long-period variations are quite irregular, more like rapidly changing
starspots than harmonic pulsations, and are likely to originate from the
companion or a contaminating star. The 7444.2-\uHz\ peak is only at
4.7\sig, but appears to be significant. We first suspected that it might be
an hitherto undescribed instrumental artefact since one other star in
Table~\ref{tbl:nonsdb} also shows its strongest high-frequency peak at
7445\,\uHz\ (but below 4\sig), and the DA+dMe star
(Section~\ref{sect:dame}, below) shows it at 5.8\sig.
However, unlike those two WDs \kic\,9822180 was observed during Q2.1,
and all the
artefacts associated with the readout cycles are double-peaked, as discussed
in Section~\ref{sect:readout}. The 7444.2-\uHz\ peak is single-peaked and
is therefore very likely to originate from the target.

\begin{figure*}
\centering
\includegraphics[height=15cm,angle=-90]{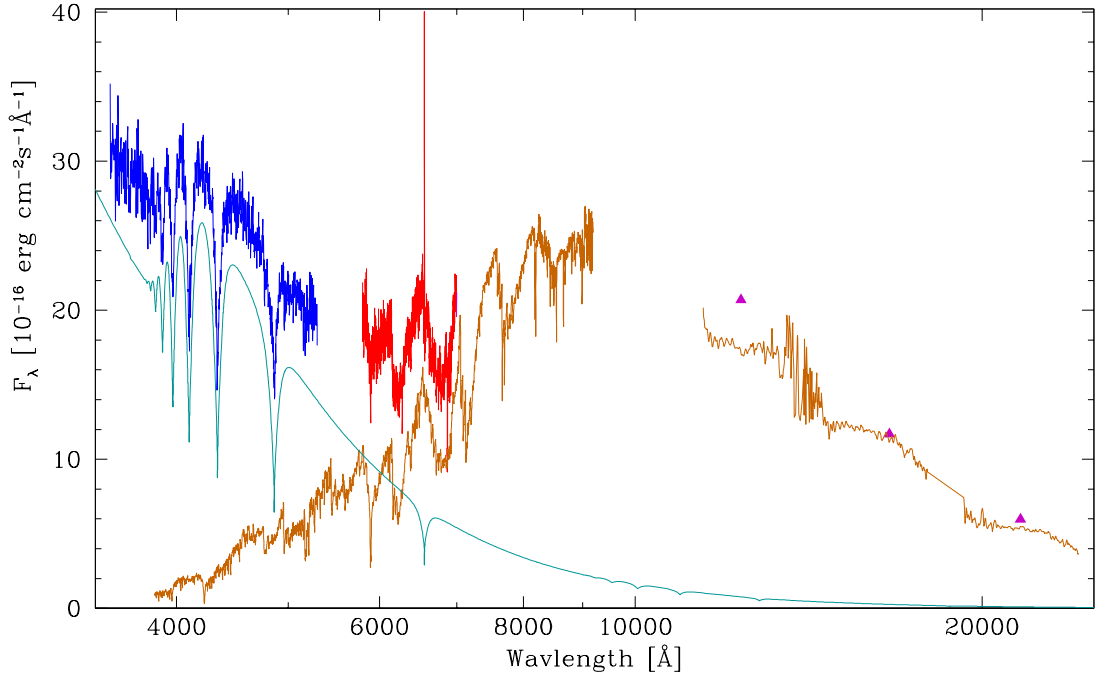}
\caption{Spectral energy distribution for J19193+5044.
The WHT/{\sc isis} spectra from the blue and red arm are shown as blue and red curves,
and the \twom\ {\em JHK} fluxes are indicated by triangles.
The lower curves show model spectra for a 20\,000\,K DA white dwarf (light blue),
and an M3 red dwarf (brown).
}
\label{fig:sed}
\end{figure*}

\subsection{Cataclysmic variables}\label{sect:cvs}

The light curves of the two objects identified as CVs from the spectroscopy
(\figref{fig:hot}) both show substantial levels of activity (top two
panels of \figref{fig:bins}).
J18561+4537 shows a light curve dominated by long-term variations.
Superimposed on these trends is a regular oscillation with a period of 0.166\,d
with a peculiar sawtooth shape. We interpret this as the orbital period
since it is typical for nova-like variables, whereas half of that period
would be exceptionally short.  In fact, J18561+4537 is remarkably
similar in terms of its spectrum, orbital period, and low-amplitude
orbital modulation to a number of known novalike variables \citep{aungwerojwit05}.
As the spectrum is disc-dominated, the large variability seen in the light curve
both on short and long time-scales must originate from the disc or
the bright spot where the gas stream hits the disc. The long period could be
ascribed to precessional behaviour of the outer disc, but could also be due
to variations in the mass-transfer rate. However, due to the strong similarity
between cycles, the latter is less likely. One can also clearly distinguish a
modulation of the orbital signal with the long-period variation, which is
consistent with precessional behaviour changing the viewing angle and therefore
the projected brightness of the hot spot. Time-resolved spectroscopy and
Doppler tomography have the potential to map the structure of the disc, and
the precessional cycle will permit several viewing angles of the system
that could reveal interesting details on accretion disc physics.

The second object, J19241+4459, is a nova-like CV of the UX\,UMa class in
or near the period gap, and it may be a new member of the rapidly growing
subclass of SW\,Sex systems \citep{williams10}.
The FT is dominated by the period at 0.122\,d, interpreted as the orbital
period, and its harmonics.

At least eleven cataclysmic variables are known to have pulsating primaries of
the ZZ\,Ceti type \citep{szkody09}. However, neither of our CVs shows any
clear signs of such pulsations. But in both systems the disc is very bright,
and the WD most likely contributes no more than $\sim$1 per cent of the flux.
All ZZ\,Ceti pulsators in CVs detected to date have been found in low
mass transfer systems, and the WDs in nova-like variables are usually
too hot for ZZ\,Ceti type pulsations. However, with the unprecedented precision
of sustained \kep\ photometry, pulsations might still be detected.
Note also that \citet{szkody10} have recently established that the instability
strip for DA pulsators in CVs extends to 15\,000\,K, substantially hotter
than for normal single ZZ\,Ceti stars.

\subsection{A DA with a flaring companion}\label{sect:dame}

The light curve shown in the bottom panel of \figref{fig:bins} is dominated
by brief flares, typical for a chromospherically active M-dwarf star.
While our original spectroscopic classification based on the {\sc isis} blue
arm spectrum shows the broad absorption
features of a single DA white dwarf, extracting the red-arm spectrum clearly
reveals the signature absorption bands of a red dwarf, and with the sharp H$\alpha$
emission feature that usually indicates chromospheric activity. The sinusoidal
modulation with a period of 3.655\,d and an amplitude of $\sim$1.1\%\
could well be the orbital period, but it might also be the rotation period of
the M-dwarf, as spots are expected to be associated with the flares.

Interestingly, after cleaning out the 3.655-d period and its harmonic, and
sigma filtering the detrended light curve to remove most of the flares, we
were able to reduce the mean noise level in the FT to 29\,ppm. A peak
at 5016.7\,\uHz\ is then found at 7.1 times the noise level. This peak
remains significant if the data is split into two halves, and processed
independently.
A second peak is also found at 7444.9\,\uHz\ at 5.8 times the noise level.
Thus, the WD primary in \kic\,12156549 might be a low level pulsator.

A calibrated version of the WHT/{\sc isis} blue+red-arm spectrum is shown
in \figref{fig:sed} together with the \twom\ IR magnitudes
($J$\,=\,12.950$\pm$0.026, $H$\,=\,12.442$\pm$0.030, $K$\,=\,12.141$\pm$0.020),
and the model spectrum of a DA white dwarf with an effective temperature
of $\sim$20\,000\,K and an M3 secondary.
The {\em JHK} colours can also be used directly to estimate the spectral
type for the companion, since the WD contribution is very low in the IR.
Following \citet{hoard07} we find that a spectral type of M4 gives the best
match, but M3 would require only a reduction in $H$--$K$ of 0.02, which is
reasonable considering the blue slope of the WD flux seen in
\figref{fig:sed}, so we conclude that the companion is most likely
of class mid M3 with M3.0 to M4.0 as a plausible range.
M3 would place the system at a distance of $\sim$140\,pc, while M4 puts it
at $\sim$100\,pc. Assuming a mass of the white dwarf of 0.6\,\msol, and
taking the $B$ magnitude to be 16.0, implies that the WD temperature
must be 24\,000\,K for 140\,pc and 18\,000\,K for 100\,pc. 24\,000\,K is at
the border of what is acceptable based on the width of the Balmer lines
in the blue spectrum, so we adopt this range as our current best estimate 
for the WD temperature, \teff\,=\,21\,000\,$\pm$\,3\,000\,K.
Given the relatively low temperature of the WD, the observed 3.65-day
modulation is unlikely to be caused by irradiation effects. A more
likely hypothesis is that it is caused by spots on the surface of the
M-dwarf, and that the 3.65-day period represents its rotation rather
than the orbital period.

Our temperature range for the WD primary shows that it is too hot to
be a regular ZZ\,Ceti pulsator, since that type of instability only occurs
inside an exclusive strip between 11\,000 and 12\,500\,K.
\citet{kurtz08} discuss candidate DA pulsators residing in the DB gap
between 30\,000 and 45\,000\,K, but excited by the same
mechanism that drives pulsations in the V777\,Her stars (or DBVs).
These stars would be structurally similar to the DB stars, but with a
thin hydrogen envelope making them appear spectroscopically as DA
white dwarfs. The temperatures of the known V777\,Her stars span 
the range between 21\,800\,K and 27\,800\,K. It is an intriguing
possibility that the DA primary in J19193+5044 could be a WD that
normally would appear as a DB star if not for a thin hydrogen
shell.
An envelope mass of between 10$^{-15}$ -- 10$^{-16}$\,\msol~is
all that is required to make a helium-atmosphere white dwarf
appear as a DA \citep{koester90}, and
the WD in J19193+5044 could accrete such an amount hydrogen
from the circumbinary envelope produced by the flaring M-dwarf.

Since the amplitudes of the pulsation peaks found in the \kep\ photometry
of \kic\,12156549 are at the 200\,ppm level, a region never before
explored for white dwarfs, it may not really make sense to confine
ourselves to the regular instability strips for known classes of
white dwarf pulsators. At these low amplitudes we may be seeing
the signature of excited pulsation modes of $\ell>2$, that are
subject to substantial geometric cancellations. Such stars may
be inhabiting temperature regions where pulsations in the readily
observable $\ell$\,=\,1,2 modes are damped, and therefore
outside the regular instability strips. Since we are exploring
uncharted territory we will refrain from further speculation until
the next cycle of \kep\ photometry can confirm the pulsations, and
our ongoing spectroscopy has established the orbit and placed
useful limits on the masses of the binary components of the system.

\section{Conclusions}

In the first four months of the survey phase (plus ten days of commissioning),
\kep\ has observed 63 compact pulsators candidates and found nine unambiguous
subdwarf\,B pulsators: seven belonging to the V1093\,Her class (long periods),
including several with hybrid behaviour, one
rapid pulsator of the V361\,Hya type also showing a single
long-period mode, indicating low-level hybrid behaviour, and one
unusual hybrid pulsator in an eclipsing sdB+dM binary. Including the
transient pulsator with a single vanishing pulsation mode means that we have
discovered ten new sdBV stars in the first half of the \kep\ survey
phase.

No clearly pulsating white dwarfs were detected in this part of the
survey, but two candidate WD pulsators with extremely low amplitudes
have been submitted for further \kep\ observations.
More intriguing is the likely detection of low amplitude pulsations that could
originate from the white dwarf component of the DA+dMe binary J19193+5044.
Being substantially hotter than the classic ZZ\,Ceti instability strip,
this object certainly merits further investigations.
The DAV candidate \kic\,10420021 and the sdOV candidate \kic\,9822180 are both
very marginal detections, at $\sim$4.5\sig, and both require further
observations before firm conclusions can be made.

If the statistics on the sdB pulsators hold up, the total number of pulsators
at the completion of the survey phase could be 18.
This number is not far from what was expected \citep{silvotti04};
a number of about ten and 2\,--\,3 was expected for the sdBV and DAV pulsators
respectively, when considering a \kep\ magnitude limit of 17.5.
However there are at least two differences. The first being that
the number of long-period sdB
pulsators is higher than expected. As pointed out in
Section~\ref{sect:photobs}, this is due to the fact that the fraction of
pulsating sdBs in the long-period instability strip is
formally 100 and not 50 per cent, as in our original conservative estimate.
The second difference is that, contrary to the expectations,
no true DAVs have been detected in this half of the sample.
Due to their intrinsic faintness, all the DA white dwarfs
observable with \kep\ belong to the galactic disc, implying that the number
of DAVs inside the \kep\ FoV is highly dependent on the magnitude limit
(while this is only marginally true for sdBs that are mostly outside the disc
at the faint end of our sample).
The number of DAVs potentially observable with \kep\ increases by a factor
four when increasing the limiting magnitude by one.
Pushing the magnitude limit to 19.5
(which would still be acceptable with respect
to the photometric accuracy), we could expect about forty DAVs in the \kep\
FoV.  In the first four months, only seven spectroscopically confirmed
DAs were observed, and in the months that remain our survey shows only six
more DAs and one single DB, leaving us cautiously optimistic that at least
one WD pulsator will be found in the final months of the survey phase.

An unexpected result is the observation that of the twelve stars in 
\figref{fig:tgplot} that lie on the EHB at temperatures
above 28\,000\,K, only two are clearly pulsating,
and only one of the remainder shows any
trace of variability. We had expected that the $\sim$10 per cent pulsator
fraction found in ground-based surveys
would increase substantially with the much extended time-base
and reduced noise level of \kep\ observations, but 25 per cent goes
some of the way.
That nine out of twelve sdB stars in the V361\,Hya instability
region do not show any trace of pulsational instability is
therefore an enduring enigma.

The detection of at least one clearly transient pulsator is a welcome
discovery, corroborating for the first time the pulsation episode discussed
by \citet{piccioni00} in HK\,Cnc (PG\,0856+121).
\citet{jeffery07} speculated that the iron-group element
enhancements, which build up due to a diffusion process over rather long
time-scales, may be disrupted by the atmospheric motions when pulsations
reach sufficient levels. They note that since $p$-modes mostly involve
vertical motion, while $g$-modes are dominated by horizontal motion,
it is possible that $p$-modes are more effective at redistributing the
iron-group elements out of the driving zone. In this way, it may be the
pulsations themselves that conspire to suppress the driving mechanism.
However, it is hard to understand how this process can act so rapidly
as we observe here, when the typical diffusion time-scales are
10$^5$\,--\,10$^6$ years.
Another possible explanation was recently suggested by \citet{theado09},
in which iron-group enhanced layers lying on top of light elements will
lead to ``iron fingers'', which are similar to the so-called ``salt fingers''
that are linked to the thermohaline convection in the oceans.
Such convection can occur on time-scales of a few thousand years,
much shorter than classical diffusion.
More work is definitely warranted both on the theoretical and observational
side in order to better understand the incidence and diversity in
pulsation power found in the subdwarf\,B pulsators.
It would be worthwhile 
to revisit the sdB stars in Table~\ref{tbl:nonpuls} later in the mission in
order to search for further transient pulsation events, and to determine
how common they are.

In this paper we have focused mostly on the many interesting sdB stars
in the compact pulsator sample. While there have been some intriguing low-level
photometric amplitudes detected in a few of the white dwarf stars in the
sample, we have not given these stars the detailed treatment they deserve.
After the final survey data have been released and processed, we will
subject the
complete set of white dwarf spectra to atmospheric analysis and discuss
them together with the photometric analysis of the second half of the
compact star sample.

Whereas the \kepmi\ has just begun, it has already lived up to its
expectations both in providing photometric data of unprecedented
quality and duration on known types of compact pulsators, as well
as revealing new and unexpected short-period phenomena.
We eagerly await the forthcoming release of the remaining half of the
survey phase, as well as the results that will come with another
order-of-magnitude increase in photometric precision that can be
reached on targets selected for study throughout the remainder
of the mission.

\section*{Acknowledgments}
The authors gratefully acknowledge the \kep\ team and everybody who
has contributed to making this mission possible.
Funding for the \kepmi\ is provided by NASA's Science Mission Directorate.

The research leading to these results has received funding from the European
Research Council under the European Community's Seventh Framework Programme
(FP7/2007--2013)/ERC grant agreement N$^{\underline{\mathrm o}}$\,227224
({\sc prosperity}), as well as from the Research Council of K.U.Leuven grant
agreement GOA/2008/04.
ACQ is supported by the Missouri Space Grant Consortium, funded by NASA.

We thank Alexander Brown (Univ.~of Colorado) for giving us early access to the
{\em Galex} GO observations of the \kep\ field.

For the spectroscopic observations presented here we acknowlede the Bok
Telescope on Kitt Peak, operated by Steward Observatory, the Nordic
Optical Telescope at the Observatorio del Roque de los Muchachos (ORM)
on La Palma, operated jointly by Denmark, Finland, Iceland,
Norway, and Sweden, and the William Herschel and Isaac Newton telescopes
also at ORM, operated by the Isaac Newton Group.

\bibliographystyle{mn2e}
\bibliography{sdbrefs}
\label{lastpage}

\end{document}

%% file: defs.tex
\newcommand{\kep}{{\em Kepler}}
\newcommand{\kepmi}{{\em Kepler Mission}}
\newcommand{\msol}{\ensuremath{\rm{M}_{\odot}}}
\newcommand{\teff}{\ensuremath{T_{\rm{eff}}}}
\newcommand{\logg}{\ensuremath{\log g}}
\newcommand{\logy}{$\log y$}

\newcommand{\lheh}{\ensuremath{\log \left(N_{\rm{He}}/N_{\rm{H}}\right)}}
\newcommand{\mkep}{{\em Kp}}
\newcommand{\corr}{\ensuremath{F_{\rm{cont}}}}
\newcommand{\fmin}{\ensuremath{f_{\rm{min}}}}
\newcommand{\fmax}{\ensuremath{f_{\rm{max}}}}
\newcommand{\fmed}{\ensuremath{f_{\rm{med}}}}
\newcommand{\porb}{\ensuremath{P_{\rm{orb}}}}
\newcommand{\uHz}{\ensuremath{\mu{\rm{Hz}}}}
\newcommand{\bd}{BD\,+42$^\circ$3250}
\newcommand{\sig}{\ensuremath{\sigma}}
\newcommand{\Fmax}{\ensuremath{f_{\rm{+}}}}
\newcommand{\Amax}{\ensuremath{A_{\rm{+}}}}
\newcommand{\kic}{{\sc kic}}
\newcommand{\twom}{{\sc 2mass}}
\newcommand{\tabkic}{\multicolumn{1}{c}{\kic}}
\newcommand{\tabsig}{\multicolumn{1}{c}{\sig}}
\newcommand{\Caii}{{Ca\,{\sc ii}}}

\newcommand{\flc}{\ensuremath{f_{\rm{lc}}}}
\newcommand{\fsc}{\ensuremath{f_{\rm{sc}}}}
\newcommand{\fnyq}{\ensuremath{f_{\rm{Nyq}}}}

\def\ie{{\sl i.e.}}

%% file: KepImn.bbl
\begin{thebibliography}{}

\bibitem[\protect\citeauthoryear{{Abell}}{{Abell}}{1966}]{abell66}
{Abell} G.~O. 1966, \apj, 144, 259

\bibitem[\protect\citeauthoryear{{Abrahamian}, {Lipovetski}, {Mickaelian} \&
  {Stepanian}}{{Abrahamian} et~al.}{1990}]{abrahamian90}
{Abrahamian} H.~V., {Lipovetski} V.~A., {Mickaelian} A.~M., {Stepanian} J.~A.
  1990, Astrofizika, 33, 213

\bibitem[\protect\citeauthoryear{{Aerts}, {Christensen-Dalsgaard} \&
  {Kurtz}}{{Aerts} et~al.}{2010}]{asteroseismology}
{Aerts} C., {Christensen-Dalsgaard} J., {Kurtz} D.~W. 2010, {Asteroseismology}.
Springer

\bibitem[\protect\citeauthoryear{{Aungwerojwit}, {G{\"a}nsicke},
  {Rodr{\'{\i}}guez-Gil}, {Hagen}, {Harlaftis}, {Papadimitriou}, {Lehto},
  {Araujo-Betancor}, {Heber}, {Fried}, {Engels} \& {Katajainen}}{{Aungwerojwit}
  et~al.}{2005}]{aungwerojwit05}
{Aungwerojwit} A., {G{\"a}nsicke} B.~T., {Rodr{\'{\i}}guez-Gil} P. et~al.,
  2005, \aap, 443, 995, arXiv:astro-ph/0507342

\bibitem[\protect\citeauthoryear{{Baran} \& {Fox Machado}}{{Baran} \& {Fox
  Machado}}{2010}]{baran10}
{Baran} A., {Fox Machado} L. 2010, \apss, in press, ...

\bibitem[\protect\citeauthoryear{{Blanchette}, {Chayer}, {Wesemael},
  {Fontaine}, {Fontaine}, {Dupuis}, {Kruk} \& {Green}}{{Blanchette}
  et~al.}{2008}]{blanchette08}
{Blanchette} J., {Chayer} P., {Wesemael} F., {Fontaine} G., {Fontaine} M.,
  {Dupuis} J., {Kruk} J.~W., {Green} E.~M. 2008, \apj, 678, 1329

\bibitem[\protect\citeauthoryear{{Bloemen}, {Marsh}, {\O stensen}, {Charpinet},
  {Degroote}, {Kawaler}, {Aerts}, {Fontaine}, {Green}, {Telting}, {Brassard}
  et~al.,}{{Bloemen} et~al.}{2010}]{bloemen10}
{Bloemen} S., {Marsh} T.~R., {\O stensen} R.~H. et~al., 2010, \mnras,
  submitted, ...

\bibitem[\protect\citeauthoryear{{Borucki}, {Koch}, {Basri}, {Batalha},
  {Brown}, {Caldwell}, {Caldwell}, {Christensen-Dalsgaard}, {Cochran},
  {DeVore}, {Dunham}, {Dupree}, {Gautier}, {Geary}, {Gilliland}
  et~al.,}{{Borucki} et~al.}{2010}]{borucki10}
{Borucki} W.~J., {Koch} D., {Basri} G. et~al., 2010, Science, 327, 977

\bibitem[\protect\citeauthoryear{{Bryson}, {Tenenbaum}, {Jenkins}, {Chandrasekaran},
  {Klaus}, {Caldwell}, {Gilliland}, {Haas}, {Dotson}, {Koch}, {Borucki}}{{Bryson}
  et~al.}{2010}]{bryson10}
{Bryson}, S.~T., {Tenenbaum}, P., {Jenkins}, J.~M. et~al., 2010, \apjl, 713,
  L97, arXiv:1001.0331

\bibitem[\protect\citeauthoryear{{Charpinet}, {Brassard}, {Fontaine}, {Green},
  {van Grootel}, {Randall} \& {Chayer}}{{Charpinet} et~al.}{2009}]{charpinet09}
{Charpinet} S., {Brassard} P., {Fontaine} G., {Green} E.~M., {van Grootel} V.,
  {Randall} S.~K., {Chayer} P. 2009, in {J.~A.~Guzik \& P.~A.~Bradley} ed.,
  American Institute of Physics Conference Series Vol.~1170 of American
  Institute of Physics Conference Series, {Progress in Sounding the Interior of
  Pulsating Hot Subdwarf Stars}.
pp 585--596

\bibitem[\protect\citeauthoryear{{Charpinet}, {Fontaine}, {Brassard}, {Chayer},
  {Rogers}, {Iglesias} \& {Dorman}}{{Charpinet} et~al.}{1997}]{charpinet97}
{Charpinet} S., {Fontaine} G., {Brassard} P., {Chayer} P., {Rogers} F.~J.,
  {Iglesias} C.~A., {Dorman} B. 1997, \apjl, 483, L123

\bibitem[\protect\citeauthoryear{{Downes}}{{Downes}}{1986}]{kpd}
{Downes} R.~A. 1986, \apjs, 61, 569

\bibitem[\protect\citeauthoryear{{Edelmann}, {Heber}, {Hagen}, {Lemke},
  {Dreizler}, {Napiwotzki} \& {Engels}}{{Edelmann} et~al.}{2003}]{edelmann03}
{Edelmann} H., {Heber} U., {Hagen} H.-J., {Lemke} M., {Dreizler} S.,
  {Napiwotzki} R., {Engels} D. 2003, \aap, 400, 939,
  arXiv:astro-ph/0301602

\bibitem[\protect\citeauthoryear{{Edelmann}, {Heber} \&
  {Napiwotzki}}{{Edelmann} et~al.}{2006}]{edelmann06}
{Edelmann} H., {Heber} U., {Napiwotzki} R. 2006, Baltic Astronomy, 15, 103

\bibitem[\protect\citeauthoryear{{Fontaine}, {Brassard}, {Charpinet} \&
  {Chayer}}{{Fontaine} et~al.}{2006}]{fontaine06}
{Fontaine} G., {Brassard} P., {Charpinet} S., {Chayer} P. 2006, Mem.
  Soc.~Astron.~Italiana, 77, 49

\bibitem[\protect\citeauthoryear{{Fontaine}, {Brassard}, {Charpinet}, {Green},
  {Chayer}, {Bill{\`e}res} \& {Randall}}{{Fontaine} et~al.}{2003}]{fontaine03}
{Fontaine} G., {Brassard} P., {Charpinet} S., {Green} E.~M., {Chayer} P.,
  {Bill{\`e}res} M., {Randall} S.~K. 2003, \apj, 597, 518

\bibitem[\protect\citeauthoryear{{For}, {Edelmann}, {Green}, {Drechsel},
  {Nesslinger} \& {Fontaine}}{{For} et~al.}{2008}]{for08}
{For} B., {Edelmann} H., {Green} E.~M., {Drechsel} H., {Nesslinger} S.,
  {Fontaine} G. 2008, in {U.~Heber, C.~S.~Jeffery, \& R.~Napiwotzki} ed., Hot
  Subdwarf Stars and Related Objects Vol.~392 of \aspcs, {KBS 13 -- a Rare
  Reflection Effect sdB Binary with an M Dwarf Secondary}.
p.~203, arXiv:0809.4517

\bibitem[\protect\citeauthoryear{{Geier}, {Heber}, {Edelmann}, {Morales-Rueda}
  \& {Napiwotzki}}{{Geier} et~al.}{2010}]{geier10a}
{Geier} S., {Heber} U., {Edelmann} H., {Morales-Rueda} L., {Napiwotzki} R.
  2010, \apss, in press, 109

\bibitem[\protect\citeauthoryear{{Gilliland}, {Brown}, {Christensen-Dalsgaard},
  {Kjeldsen}, {Aerts}, {Appourchaux}, {Basu}, {Bedding}, {Chaplin}, {Cunha},
  {De Cat}, {De Ridder}, {Guzik}, {Handler} et~al.,}{{Gilliland}
  et~al.}{2010a}]{gilliland10a}
{Gilliland} R.~L., {Brown} T.~M., {Christensen-Dalsgaard} J. et~al., 2010,
  \pasp, 122, 131,
  arXiv:1001.0139

\bibitem[\protect\citeauthoryear{{Gilliland}, {Jenkins}, {Borucki}, {Bryson},
  {Caldwell}, {Clarke}, {Dotson}, {Haas}, {Hall}, {Klaus}, {Koch}, {McCauliff},
  {Quintana}, {Twicken} \& {van Cleve}}{{Gilliland}
  et~al.}{2010b}]{gilliland10b}
{Gilliland} R.~L., {Jenkins} J.~M., {Borucki} W.~J. et~al., 2010, \apjl, 713,
  L160, arXiv:1001.0142 

\bibitem[\protect\citeauthoryear{{Green}, {Fontaine}, {Reed}, {Callerame},
  {Seitenzahl}, {White}, {Hyde}, {{\O}stensen}, {Cordes}, {Brassard}, {Falter},
  {Jeffery}, {Dreizler}, {Schuh}, {Giovanni}, {Edelmann}, {Rigby} \&
  {Bronowska}}{{Green} et~al.}{2003}]{green03}
{Green} E.~M., {Fontaine} G., {Reed} M.~D. et~al., 2003, \apjl, 583, L31,
   arXiv:astro-ph/0210285 

\bibitem[\protect\citeauthoryear{{Heber}}{{Heber}}{2009}]{heber09}
{Heber} U. 2009, \araa, 47, 211

\bibitem[\protect\citeauthoryear{{Heber}, {Reid} \& {Werner}}{{Heber}
  et~al.}{2000}]{heber00}
{Heber} U., {Reid} I.~N., {Werner} K. 2000, \aap, 363, 198

\bibitem[\protect\citeauthoryear{{Hoard}, {Wachter}, {Sturch}, {Widhalm},
  {Weiler}, {Pretorius}, {Wellhouse} \& {Gibiansky}}{{Hoard}
  et~al.}{2007}]{hoard07}
{Hoard} D.~W., {Wachter} S., {Sturch} L.~K., {Widhalm} A.~M., {Weiler} K.~P.,
  {Pretorius} M.~L., {Wellhouse} J.~W., {Gibiansky} M. 2007, \aj, 134, 26,
  arXiv:astro-ph/0702754 

\bibitem[\protect\citeauthoryear{{Jeffery} \& {Saio}}{{Jeffery} \&
  {Saio}}{2007}]{jeffery07}
{Jeffery} C.~S., {Saio} H. 2007, \mnras, 378, 379

\bibitem[\protect\citeauthoryear{{Kawaler}, {Bond}, {Sherbert} \&
  {Watson}}{{Kawaler} et~al.}{1994}]{kawaler93}
{Kawaler} S.~D., {Bond} H.~E., {Sherbert} L.~E., {Watson} T.~K. 1994, \aj, 107,
  298

\bibitem[\protect\citeauthoryear{{Kawaler}, {Reed}, {Quint}, {\O stensen},
  {Silvotti}, {Baran}, {Charpinet}, {Bloemen}, {Kurtz}, {Telting}, {Handler}
  et~al.,}{{Kawaler} et~al.}{2010a}]{kawaler10a}
{Kawaler} S.~D., {Reed} M., {Quint} A.~M. et~al., 2010a, \mnras, submitted

\bibitem[\protect\citeauthoryear{{Kawaler}, {Reed}, {\O stensen}, {Bloemen},
  {Kurtz}, {Quint}, {Silvotti}, {Baran}, {Green}, {Charpinet}, {Telting},
  {Aerts}, {Handler} et~al.,}{{Kawaler} et~al.}{2010b}]{kawaler10b}
{Kawaler} S.~D., {Reed} M., {\O stensen} R.~H. et~al., 2010b, \mnras, submitted


\bibitem[\protect\citeauthoryear{{Kilkenny}}{{Kilkenny}}{2010}]{kilkenny10a}
{Kilkenny} D. 2010, \apss, in press, 82

\bibitem[\protect\citeauthoryear{{Kilkenny}, {Fontaine}, {Green} \&
  {Schuh}}{{Kilkenny} et~al.}{2010}]{kilkenny10b}
{Kilkenny} D., {Fontaine} G., {Green} E.~M., {Schuh} S. 2010, Information
  Bulletin on Variable Stars, 5927, 1

\bibitem[\protect\citeauthoryear{{Kilkenny}, {Koen}, {O'Donoghue} \&
  {Stobie}}{{Kilkenny} et~al.}{1997}]{kilkenny97}
{Kilkenny} D., {Koen} C., {O'Donoghue} D., {Stobie} R.~S. 1997, \mnras, 285,
  640

\bibitem[\protect\citeauthoryear{{Koester} \& {Chanmugam}}{{Koester} \&
  {Chanmugam}}{1990}]{koester90}
{Koester} D., {Chanmugam} G. 1990, Reports on Progress in Physics, 53, 837

\bibitem[\protect\citeauthoryear{{Kronberger}, {Teutsch}, {Alessi}, {Steine},
  {Ferrero}, {Graczewski}, {Juchert}, {Patchick}, {Riddle}, {Saloranta},
  {Schoenball} \& {Watson}}{{Kronberger} et~al.}{2006}]{kronberger06}
{Kronberger} M., {Teutsch} P., {Alessi} B. et~al., 2006, \aap, 447, 921,
   arXiv:gr-qc/0511021

\bibitem[\protect\citeauthoryear{{Kurtz}, {Shibahashi}, {Dhillon}, {Marsh} \&
  {Littlefair}}{{Kurtz} et~al.}{2008}]{kurtz08}
{Kurtz} D.~W., {Shibahashi} H., {Dhillon} V.~S., {Marsh} T.~R., {Littlefair}
  S.~P. 2008, \mnras, 389, 1771

\bibitem[\protect\citeauthoryear{{Lutz}, {Schuh}, {Silvotti}, {Bernabei},
  {Dreizler}, {Stahn} \& {H{\"u}gelmeyer}}{{Lutz} et~al.}{2009}]{lutz09}
{Lutz} R., {Schuh} S., {Silvotti} R., {Bernabei} S., {Dreizler} S., {Stahn} T.,
  {H{\"u}gelmeyer} S.~D. 2009, \aap, 496, 469,
  arXiv:0901.4523

\bibitem[\protect\citeauthoryear{{Marsh}, {Barstow}, {Buckley}, {Burleigh},
  {Holberg}, {Koester}, {O'Donoghue}, {Penny} \& {Sansom}}{{Marsh}
  et~al.}{1997}]{marsh97}
{Marsh} M.~C., {Barstow} M.~A., {Buckley} D.~A. et~al., 1997, \mnras, 286, 369

\bibitem[\protect\citeauthoryear{{Martin}, {Fanson}, {Schiminovich},
  {Morrissey}, {Friedman}, {Barlow}, {Conrow}, {Grange}, {Jelinsky},
  {Milliard}, {Siegmund}, {Bianchi}, {Byun} et~al.,}{{Martin}
  et~al.}{2005}]{GALEX}
{Martin} D.~C., {Fanson} J., {Schiminovich} D. et~al., 2005, \apjl, 619, L1,
   arXiv:astro-ph/0411302

\bibitem[\protect\citeauthoryear{{Montgomery}, {Williams}, {Winget}, {Dufour},
  {De Gennaro} \& {Liebert}}{{Montgomery} et~al.}{2008}]{montgomery08}
{Montgomery} M.~H., {Williams} K.~A., {Winget} D.~E., {Dufour} P., {De Gennaro}
  S., {Liebert} J. 2008, \apjl, 678, L51,
   arXiv:0803.2646

\bibitem[\protect\citeauthoryear{{Morales-Rueda}, {Maxted}, {Marsh}, {North} \&
  {Heber}}{{Morales-Rueda} et~al.}{2003}]{morales-rueda03}
{Morales-Rueda} L., {Maxted} P.~F.~L., {Marsh} T.~R., {North} R.~C., {Heber} U.
  2003, \mnras, 338, 752,
  arXiv:astro-ph/0209472

\bibitem[\protect\citeauthoryear{{Napiwotzki}}{{Napiwotzki}}{1997}]{napiwotzki%
97}
{Napiwotzki} R. 1997, \aap, 322, 256

\bibitem[\protect\citeauthoryear{{Napiwotzki}}{{Napiwotzki}}{1999}]{napiwotzki%
99}
{Napiwotzki} R. 1999, \aap, 350, 101,
arXiv:astro-ph/9908181

\bibitem[\protect\citeauthoryear{{Oreiro}, {P{\'e}rez Hern{\'a}ndez}, {Ulla},
  {Garrido}, {{\O}stensen} \& {MacDonald}}{{Oreiro} et~al.}{2005}]{oreiro05}
{Oreiro} R., {P{\'e}rez Hern{\'a}ndez} F., {Ulla} A., {Garrido} R.,
  {{\O}stensen} R., {MacDonald} J. 2005, \aap, 438, 257

\bibitem[\protect\citeauthoryear{{Oreiro}, {Ulla}, {P{\'e}rez Hern{\'a}ndez},
  {{\O}stensen}, {Rodr{\'{\i}}guez L{\'o}pez} \& {MacDonald}}{{Oreiro}
  et~al.}{2004}]{oreiro04}
{Oreiro} R., {Ulla} A., {P{\'e}rez Hern{\'a}ndez} F., {{\O}stensen} R.,
  {Rodr{\'{\i}}guez L{\'o}pez} C., {MacDonald} J. 2004, \aap, 418, 243

\bibitem[\protect\citeauthoryear{{\O stensen}}{{\O stensen}}{2009}]{ostensen09}
{\O stensen} R. 2009, Communications in Asteroseismology, 159, 75
   arXiv:0901.1618

\bibitem[\protect\citeauthoryear{{{\O}stensen}, {Heber}, {Silvotti}, {Solheim},
  {Dreizler} \& {Edelmann}}{{{\O}stensen} et~al.}{2001a}]{ostensen01b}
{{\O}stensen} R., {Heber} U., {Silvotti} R., {Solheim} J.-E., {Dreizler} S.,
  {Edelmann} H. 2001, \aap, 378, 466

\bibitem[\protect\citeauthoryear{{{\O}stensen}, {Solheim}, {Heber}, {Silvotti},
  {Dreizler} \& {Edelmann}}{{{\O}stensen} et~al.}{2001b}]{ostensen01a}
{{\O}stensen} R., {Solheim} J.-E., {Heber} U., {Silvotti} R., {Dreizler} S.,
  {Edelmann} H. 2001, \aap, 368, 175

\bibitem[\protect\citeauthoryear{{\O stensen}, {Oreiro}, {Solheim}, {Heber},
  {Silvotti}, {Gonz{\'a}lez-P{\'e}rez}, {Ulla}, {P{\'e}rez Hern{\'a}ndez},
  {Rodr{\'{\i}}guez-L{\'o}pez} \& {Telting}}{{\O stensen}
  et~al.}{2010a}]{sdbnot}
{\O stensen} R.~H., {Oreiro} R., {Solheim} J. et~al., 2010a, \aap, 513, A6,
  arXiv:1001.3657

\bibitem[\protect\citeauthoryear{{\O stensen}, {Green}, {Bloemen}, {Marsh},
  {Oreiro}, {Laird}, {Morris}, {Moriyama}, {Reed}, {Kawaler}, {Aerts},
  {Vu\v{c}kovi\'{c}}, {Degroote}, {Telting} et~al.,}{{\O stensen}
  et~al.}{2010b}]{2m1938}
{\O stensen} R.~H., {Green} E.~M., {Bloemen} S. et~al., 2010b, \mnras,
  submitted, ...

\bibitem[\protect\citeauthoryear{{O'Toole} \& {Heber}}{{O'Toole} \&
  {Heber}}{2006}]{otoole06}
{O'Toole} S.~J., {Heber} U. 2006, \aap, 452, 579,
  arXiv:astro-ph/0603069 

\bibitem[\protect\citeauthoryear{{Piccioni}, {Bartolini}, {Bernabei}, {Bruni},
  {Galleti}, {Giovannelli}, {Guarnieri}, {Sabau-Graziati}, {Silvotti}, {Ulla}
  \& {Valentini}}{{Piccioni} et~al.}{2000}]{piccioni00}
{Piccioni} A., {Bartolini} C., {Bernabei} S. et~al., 2000, \aap, 354, L13

\bibitem[\protect\citeauthoryear{{Podsiadlowski}, {Han}, {Lynas-Gray} \&
  {Brown}}{{Podsiadlowski} et~al.}{2008}]{podsi08}
{Podsiadlowski} P., {Han} Z., {Lynas-Gray} A.~E., {Brown} D. 2008, in
  {U.~Heber, C.~S.~Jeffery, \& R.~Napiwotzki} ed., Hot Subdwarf Stars and
  Related Objects Vol.~392 of \aspcs, {Hot Subdwarfs in Binaries as the Source
  of the Far-UV Excess in Elliptical Galaxies}.
p.~15, arXiv:0808.0574

\bibitem[\protect\citeauthoryear{{Ramsay}, {Napiwotzki}, {Hakala} \&
  {Lehto}}{{Ramsay} et~al.}{2006}]{ramsay06}
{Ramsay} G., {Napiwotzki} R., {Hakala} P., {Lehto} H. 2006, \mnras, 371, 957,
   arXiv:astro-ph/0606628

\bibitem[\protect\citeauthoryear{{Randall}, {Green}, {Fontaine}, {Brassard},
  {Terndrup}, {Brown}, {Fontaine}, {Zacharias} \& {Chayer}}{{Randall}
  et~al.}{2006}]{randall06}
{Randall} S.~K., {Green} E.~M., {Fontaine} G. et~al., 2006, \apj, 645, 1464

\bibitem[\protect\citeauthoryear{{Reed}, {Kawaler}, {\O stensen}, {Bloemen},
  {Baran}, {Silvotti}, {Charpinet}, {Quint}, {Handler}, {Gilliland}
  et~al.,}{{Reed} et~al.}{2010}]{reed10a}
{Reed} M., {Kawaler} S.~D., {\O stensen} R.~H. et~al., 2010, \mnras, submitted,
  ...

\bibitem[\protect\citeauthoryear{{Reed}, {Eggen}, {Zhou}, {Terndrup}, {Harms},
  {An} \& {Hashier}}{{Reed} et~al.}{2006}]{reed06}
{Reed} M.~D., {Eggen} J.~R., {Zhou} A.-Y., {Terndrup} D.~M., {Harms} S.~L.,
  {An} D., {Hashier} M.~A. 2006, \mnras, 369, 1529,
   arXiv:astro-ph/0603804

\bibitem[\protect\citeauthoryear{{Schuh}, {Huber}, {Dreizler}, {Heber},
  {O'Toole}, {Green} \& {Fontaine}}{{Schuh} et~al.}{2006}]{schuh06}
{Schuh} S., {Huber} J., {Dreizler} S., {Heber} U., {O'Toole} S.~J., {Green}
  E.~M., {Fontaine} G. 2006, \aap, 445, L31,
  arXiv:astro-ph/0510831

\bibitem[\protect\citeauthoryear{{Silvotti}}{{Silvotti}}{2004}]{silvotti04}
{Silvotti} R. 2004, in {F.~Favata, S.~Aigrain, \& A.~Wilson} ed., Stellar
  Structure and Habitable Planet Finding Vol.~538 of ESA Special Publication,
  {Asteroseismology of white dwarfs and subdwarf B stars}.
p.~141

\bibitem[\protect\citeauthoryear{{Silvotti}, {Bonanno}, {Bernabei}, {Fontaine},
  {Charpinet}, {Leccia}, {Kjeldsen}, {Janulis}, {Frasca}, {{\O}stensen}, {Kim},
  {Park}, {Jiang}, {Reed}, {Patterson}, {Gietzen} et~al.,}{{Silvotti}
  et~al.}{2006}]{silvotti06}
{Silvotti} R., {Bonanno} A., {Bernabei} S. et~al., 2006, \aap, 459, 557

\bibitem[\protect\citeauthoryear{{Silvotti}, {Handler}, {Schuh}, {Castanheira}
  \& {Kjeldsen}}{{Silvotti} et~al.}{2009}]{silvotti09}
{Silvotti} R., {Handler} G., {Schuh} S., {Castanheira} B., {Kjeldsen} H. 2009,
  Communications in Asteroseismology, 159, 97,
  arXiv:0901.1011

\bibitem[\protect\citeauthoryear{{Silvotti}, {{\O}stensen}, {Heber}, {Solheim},
  {Dreizler} \& {Altmann}}{{Silvotti} et~al.}{2002}]{silvotti02a}
{Silvotti} R., {{\O}stensen} R., {Heber} U., {Solheim} J.-E., {Dreizler} S.,
  {Altmann} M. 2002, \aap, 383, 239

\bibitem[\protect\citeauthoryear{{Silvotti}, {Schuh}, {Janulis}, {Solheim},
  {Bernabei}, {{\O}stensen}, {Oswalt}, {Bruni}, {Gualandi}, {Bonanno},
  {Vauclair}, {Reed}, {Chen}, {Leibowitz}, {Paparo} et~al.,}{{Silvotti}
  et~al.}{2007}]{silvotti07}
{Silvotti} R., {Schuh} S., {Janulis} R. et~al., 2007, \nat, 449, 189

\bibitem[\protect\citeauthoryear{{Skrutskie}, {Cutri}, {Stiening}, {Weinberg},
  {Schneider}, {Carpenter}, {Beichman}, {Capps}, {Chester}, {Elias}, {Huchra},
  {Liebert}, {Lonsdale}, {Monet} et~al.,}{{Skrutskie} et~al.}{2006}]{twomass}
{Skrutskie} M.~F., {Cutri} R.~M., {Stiening} R. et~al., 2006, \aj, 131, 1163

\bibitem[\protect\citeauthoryear{{Stoughton}, {Lupton}, {Bernardi}, {Blanton},
  {Burles}, {Castander}, {Connolly}, {Eisenstein}, {Frieman}, {Hennessy},
  {Hindsley}, {Ivezi{\'c}}, {Kent}, {Kunszt}, {Lee}, {Meiksin}, {Munn}
  et~al.,}{{Stoughton} et~al.}{2002}]{SDSS}
{Stoughton} C., {Lupton} R.~H., {Bernardi} M. et~al., 2002, \aj, 123, 485

\bibitem[\protect\citeauthoryear{{Szkody}, {Mukadam}, {G{\"a}nsicke}, {Henden},
  {Templeton}, {Holtzman}, {Montgomery}, {Howell}, {Nitta}, {Sion}, {Schwartz}
  \& {Dillon}}{{Szkody} et~al.}{2010}]{szkody10}
{Szkody} P., {Mukadam} A., {G{\"a}nsicke} B.~T. et~al., 2010, \apj, 710, 64,
   arXiv:1001.0192

\bibitem[\protect\citeauthoryear{{Szkody}, {Mukadam}, {G\"ansicke}, {Henden},
  {Nitta}, {Sion} \& {Townsley}}{{Szkody} et~al.}{2009}]{szkody09}
{Szkody} P., {Mukadam} A.~S., {G\"ansicke} B.~T., {Henden} A., {Nitta} A.,
  {Sion} E.~M., {Townsley} D. 2009, in {S.~J.~Murphy \& M.~S.~Bessell} ed., The
  Eighth Pacific Rim Conference on Stellar Astrophysics Vol.~404 of \aspcs,
  {The Accreting, Pulsating White Dwarfs in Cataclysmic Variables}.
p.~229

\bibitem[\protect\citeauthoryear{{Th{\'e}ado}, {Vauclair}, {Alecian} \& {Le
  Blanc}}{{Th{\'e}ado} et~al.}{2009}]{theado09}
{Th{\'e}ado} S., {Vauclair} S., {Alecian} G., {Le Blanc} F. 2009, \apj, 704,
  1262, arXiv:0908.1534

\bibitem[\protect\citeauthoryear{{Van Grootel}, {Charpinet}, {Fontaine},
  {Brassard}, {Green}, {Randall}, {Silvotti}, {\O stensen}, {Gilliland}
  et~al.,}{{Van Grootel} et~al.}{2010}]{VanGrootel10}
{Van Grootel} V., {Charpinet} S., {Fontaine} G. et~al., 2010, \apj, 718, L97

\bibitem[\protect\citeauthoryear{{Williams}, {de Martino}, {Silvotti}, {Bruni},
  {Dufour}, {Riecken}, {Kronberg}, {Mukadam} \& {Handler}}{{Williams}
  et~al.}{2010}]{williams10}
{Williams} K.~A., {de Martino} D., {Silvotti} R. et~al., 2010, \aj, 139, 2587,
  arXiv:1004.3743 

\bibitem[\protect\citeauthoryear{{Woudt}, {Kilkenny}, {Zietsman}, {Warner},
  {Loaring}, {Copley}, {Kniazev}, {V{\"a}is{\"a}nen}, {Still}, {Stobie},
  {Burgh}, {Nordsieck}, {Percival}, {O'Donoghue} \& {Buckley}}{{Woudt}
  et~al.}{2006}]{woudt06}
{Woudt} P.~A., {Kilkenny} D., {Zietsman} E. et~al., 2006, \mnras, 371, 1497,
  arXiv:astro-ph/0607171

\bibitem[\protect\citeauthoryear{{Yanny}, {Rockosi}, {Newberg}, {Knapp},
  {Adelman-McCarthy}, {Alcorn}, {Allam}, {Allende Prieto}, {An}, {Anderson},
  {Anderson}, {Bailer-Jones}, {Bastian}, {Beers}, {Bell}, {Belokurov},
  {Bizyaev} et~al.,}{{Yanny} et~al.}{2009}]{SEGUE}
{Yanny} B., {Rockosi} C., {Newberg} H.~J. et~al., 2009, \aj, 137, 4377,
  arXiv:0902.1781

\end{thebibliography}
